\documentclass[iop]{emulateapj}

\usepackage{paralist}
\usepackage{graphics,graphicx}
\newcommand{\nar}{New Astronomy Reviews}

\begin{document}

\shortauthors{Mineo, Rappaport, Steinhorn, Levine, Gilfanov \& Pooley}
\shorttitle{ULXs population of colliding Galaxies NGC 2207/IC 2163}
\title{Spatially resolved star formation image and the ULX population in NGC2207/IC2163}

\author{S. Mineo\altaffilmark{1}, S. Rappaport\altaffilmark{2,6}, B. Steinhorn\altaffilmark{3}, A. Levine\altaffilmark{4}, M. Gilfanov\altaffilmark{5}, \& D. Pooley\altaffilmark{6}}

\altaffiltext{1}{Harvard-Smithsonian Center for Astrophysics, 60 Garden Street Cambridge, MA 02138 USA; smineo@head.cfa.harvard.edu} 
\altaffiltext{2}{37-602B, M.I.T. Department of Physics and Kavli
 Institute for Astrophysics and Space Research, 70 Vassar St.,
 Cambridge, MA, 02139; sar@mit.edu} 
  \altaffiltext{3}{Harvard-MIT Division of Health Sciences and Technology, 
Harvard Medical School, 260 Longwood Ave., Boston, MA 02115; bsteinho@mit.edu} 
 \altaffiltext{4}{M.I.T. Kavli
 Institute for Astrophysics and Space Research, Room 37-575, 70 Vassar St.,
 Cambridge, MA, 02139; aml@space.mit.edu} 
\altaffiltext{5}{Max Planck Institut f\"ur Astrophysik, Karl-Schwarzschild-Str. 1 85741 Garching, Germany; Space Research Institute of Russian Academy of Sciences, Profsoyuznaya 84/32 117997 Moscow, Russia; gilfanov@mpa-garching.mpg.de} 
\altaffiltext{6}{Eureka Scientific, 5248 Valley View Road,
El Sobrante, CA 94803-3435; pooley@gmail.com}

\begin{abstract}

The colliding galaxy pair NGC 2207/IC 2163, at a distance of $\sim$39 Mpc, was observed with {\em Chandra}, and an analysis reveals 28 well resolved X-ray sources, including 21 ultraluminous X-ray sources (ULXs) with $L_{\rm{X}}\gtrsim 10^{39}$ erg s$^{-1}$, as well as the nucleus of NGC~2207. The number of ULXs is comparable with the largest numbers of ULXs per unit mass in any galaxy yet reported.  In this paper we report on these sources, and quantify how their locations correlate with the local star formation rates seen in spatially-resolved star formation rate density images that we have constructed using combinations of {\em Galex} FUV and {\em Spitzer} 24$\mu$m images.  We show that the {\em numbers} of ULXs are strongly correlated with the local star formation rate densities surrounding the sources, but that the {\em luminosities} of these sources are not strongly correlated with star formation rate density.
\end{abstract}


\section{Introduction}
\label{sec:intro}
Galaxy collisions can provide information on (i) galaxy dynamics, (ii)
triggers of star formation, and (iii) the origins of ultraluminous
X-ray sources, i.e., off-nuclear X-ray point sources whose luminosity exceeds the maximum isotropic emission expected from a $\sim$$10\,M_{\odot}$ black hole (BH): $L_{\rm{X}}\gtrsim 10^{39}$ erg
s$^{-1}$ \citep[hereafter, ``ULXs'', see review
by][]{2011NewAR..55..166F}.  When one or both of the colliding
galaxies has a high gas content, spectacular bursts of star formation
may be triggered. Some fraction of the more massive stars ($\gtrsim 10\,M_{\odot}$) that happen
to be formed in binary systems evolve to become high-mass
X-ray binaries (HMXBs) with neutron-star (NS) or stellar-mass BH accretors. Substantial numbers
of ULXs may also be produced. The Antennae \citep{1995AJ....109..960W, 2002ApJ...577..726Z}, the Cartwheel \citep{1995ApJ...455..524H, 2003ApJ...596L.171G, 2006MNRAS.373.1627W},
and Arp 147 \citep{2010ApJ...721.1348R} grandly illustrate these phenomena. Although the accepted assumption was that HMXBs would be Eddington limited ($L_{\rm{X}}\lesssim 10^{39}$ erg s$^{-1}$), there is a growing body of evidence linking ULXs to these sources \citep[e.g.,][]{2003MNRAS.339..793G, 2005Sci...307..533F,2012MNRAS.419.2095M}.

Compelling simulations of colliding galaxies have been carried out 
\citep[see, e.g.,][]{1976ApJ...209..382L, 1978IAUS...79..109T, 1992ApJ...399L..51G,1994ApJ...431L...9M,1996ApJ...468..532A,1997ApJ...474..686H, 2005MNRAS.364...69S}, and have produced examples that are remarkably similar in appearance to the most tidally disturbed collisional pairs.

The ULXs seen in these galaxies are of special interest, especially those with $L_{\rm{X}}\gtrsim 10^{40}$ erg s$^{-1}$. Assuming that these objects are BHs accreting at the Eddington limit, their luminosity gives a limit on the mass of the accreting BH of $\sim$$50-100\,M_{\odot}$. These masses are substantially higher than those of Galactic BHs \citep[$M \lesssim 20\,M_{\odot}$][]{2010ApJ...725..1918O} and could be related with the metallicity of the host galaxy \citep[e.g.,][]{2010ApJ...714.1217B,2010MNRAS.408..234M}. The population of ULXs may therefore consist of sources of a different nature, depending on their luminosity. Sources with $L_{\rm{X}} \lesssim 10^{40}$ erg s$^{-1}$ might be powered by accreting BHs of mass $3\,M_{\odot} \lesssim M \lesssim 100\,M_{\odot}$, whereas the nature of the brightest sources ($L_{\rm{X}}\gtrsim 10^{40}$ erg s$^{-1}$) still represents an enigma. They might be $\sim$$10-20\,M_{\odot}$ BHs emitting at $\sim$$5$ times the Eddington luminosity, BHs with masses $\sim$$100\,M_{\odot}$ emitting at the Eddington limit, or the so-called ``intermediate-mass black holes'' \citep[``IMBHs'' $10^{2}\,M_{\odot} \lesssim M \lesssim 10^{5}\,M_{\odot}$; see, e.g.,][]{1999ApJ...519...89C} emitting below the Eddington limit, or some combination thereof. At the highest end of the ULX luminosity function (approaching $\sim$$10^{41}$ erg sec$^{-1}$) it becomes increasingly difficult to see how the requisite luminosity, even if somewhat beamed, could be radiated near a stellar-mass black hole \citep[see, e.g.,][]{2008ApJ...688.1235M}. The maximum observed luminosity for such sources is $\sim$$10^{42}$ erg sec$^{-1}$ \citep{2009Natur.460...73F}.

The IMBHs are of extreme importance as they are thought to be the building blocks of super-massive BHs \citep[``SMBHs'' $M\sim 10^{6}\,M_{\odot}$; e.g.,][]{2010Natur.466.1049V}. However, a conceptual problem with IMBH accretors is their formation and their subsequent capture of a massive donor star.  \citet{2004Natur.428..724P}, 
among others, have proposed that runaway star collisions in newly formed 
massive star clusters lead to the formation of supermassive stars 
(e.g., $\gtrsim 500~M_\odot$) which, in turn, evolve to form IMBHs.  
Theoretical problems with this scenario include the highly uncertain 
evolution of supermassive stars, and an implausibly high efficiency for 
producing the requisite numbers of IMBHs \citep[see][]{2004MNRAS.347L..18K}.

Previous studies have shown a correlation between the {\em overall} star
formation rate (hereafter ``SFR'') in a galaxy and the number of
luminous X-ray sources \citep[e.g.,][]{2003MNRAS.339..793G,
2010MNRAS.408..234M, 2011ApJ...741...49S, 2012MNRAS.419.2095M, 2012AJ....143..144S}.  The
SFRs may be estimated using any of a wide variety of indicators: ultraviolet (UV) continuum, recombination lines (H$\alpha$), forbidden lines ($[OII]$), far-infrared (FIR) continuum, thermal radio emission luminosities individually, or in combination \citep[see, e.g.,][for a review]{1998ARA&A..36..189K}.  \citet{2008AJ....136.2782L} suggest that a linear combination of the {\em Galex} FUV and {\em Spitzer} 24 $\mu$m bands
(see their eqs.\,D10 and D11) is particularly good in this respect. It is
possible to follow Leroy et al.~and utilize this prescription to
produce complete SFR {\em images} of galaxies with a few arcsec resolution.

In this paper we introduce a new technique to investigate the relation between the number and luminosity of bright X-ray point sources and the {\em local} surrounding SFR. This allows us to probe these relations even in the case of small numbers of luminous X-ray sources. The analysis involves a direct quantitative comparison of the spatial structures in these SFR 
images with the {\em Chandra} X-ray images. In addition,
as has been discussed (Calzetti et al.~2007; Rappaport et al.~2012, in preparation; Mark Krumholz, private communication 2012), the {\em Galex} intensities
tend to indicate the somewhat older regions of star formation ($\gtrsim 30$ Myr, after
the obscuring dust has already been cleared), while the {\em Spitzer} 24 $\mu$m 
images reveal younger star formation (i.e., $5-10$ Myr, and still dust enshrouded)
which may be more closely related to the upper end of the ULX luminosities (i.e., with $L_{\rm{X}}\gtrsim 10^{40}$ erg sec$^{-1}$).  (See also the closely related work of \citet{2009ApJ...703..159S, 2010AJ....139.1066Y, 2012AJ....144..156K}). Thus, some of the theoretical ideas concerning the formation and evolution of very massive binaries can be investigated.

\begin{figure*}
\begin{center}
\includegraphics[width=1.0\linewidth]{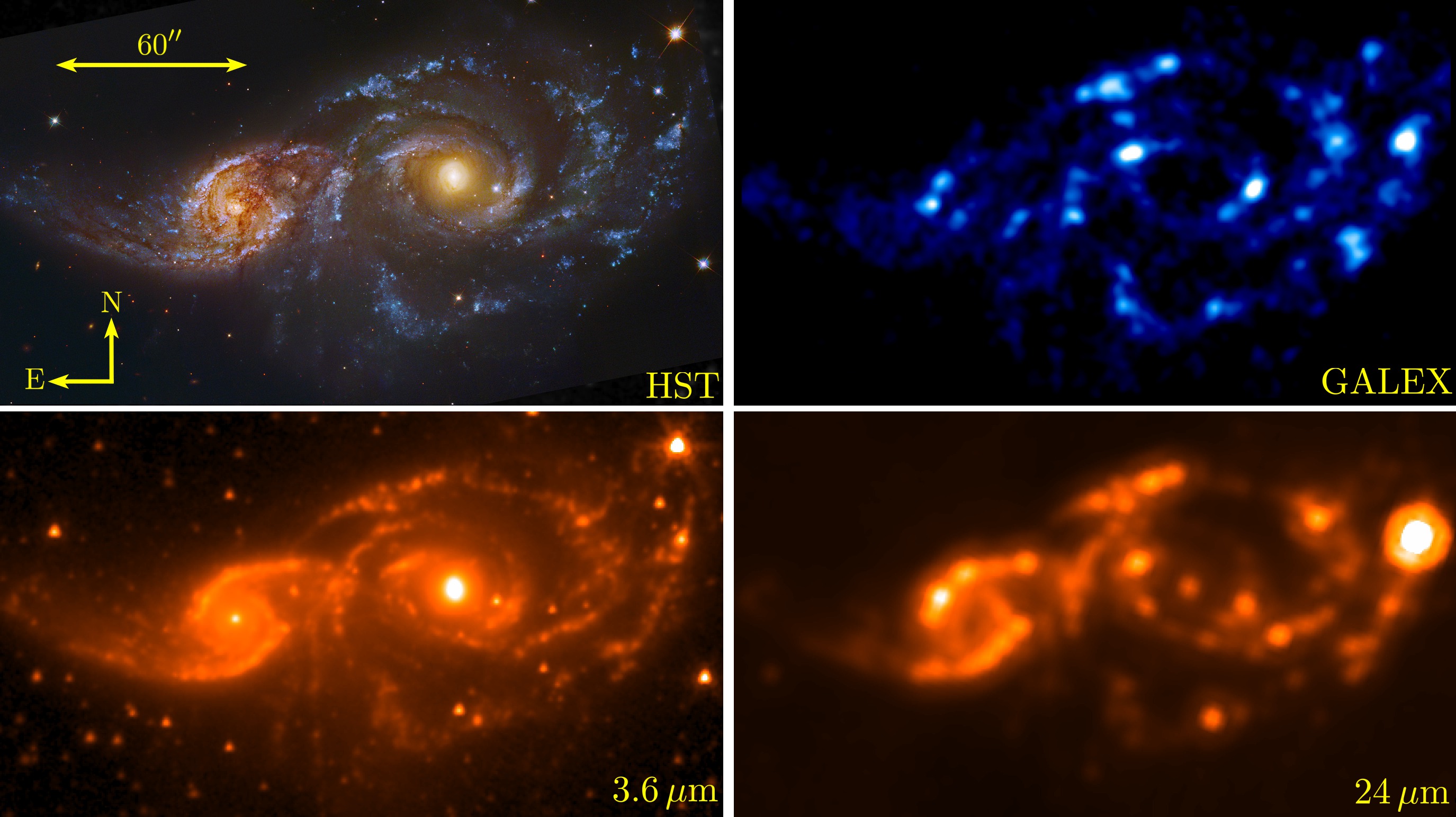}
\caption{Montage of images of NGC~2207/IC~2163 taken with {\em HST}, {\em
Galex} (FUV), and {\em Spitzer} (3.6 and 24 $\mu$m).  The distance to these galaxies is $\sim$39 Mpc.}
\label{fig:montage}
\end{center}
\end{figure*}

We report here on our analysis of the relation between the location of luminous 
X-ray sources and the star formation rate density in the colliding galaxy pair 
NGC 2207 \& IC 2163 \citep[see, e.g.,][]{2012AJ....144..156K}. We utilize archival 
{\em Chandra}, {\em Galex}, {\em Spitzer} and Two Micron All Sky Survey (2MASS) images of these galaxies. The galaxy pair is at an estimated distance of $39.6^{+5.9}_{-5.1}$ Mpc \citep{1982ApJ...254....1A}. This is the redshift-independent distance with the smallest uncertainty provided by the NASA/IPAC Extragalactic Database (NED)\footnote{http://ned.ipac.caltech.edu/}, and it is based on SN-Ia measurements. A montage of images of NGC 2207/IC 2163 taken with {\em HST}, {\em Galex}, and {\em Spitzer} is shown in Fig.~\ref{fig:montage} (NGC 2207 is the larger galaxy on the right).  Our analysis of the archival {\em Chandra} data reveal a total of 22 X-ray point sources with $L_{\rm{X}}\gtrsim 10^{39}$ erg s$^{-1}$ within the $D25$ ellipse of NGC~2207 \& IC~2163 \citep{1991trcb.book.....D}. Such a production efficiency of luminous X-ray sources per unit stellar mass is comparable with that of the Antennae pair of colliding galaxies (see Sect.~\ref{sec:int_sfr} and Table~\ref{tab:sfr}).  

Another advantage of studying the colliding galaxy pair NGC 2207/IC2163, in addition to the fact that it is relatively close and well studied in numerous wavebands, is that the dynamics of its collision have been extensively modeled \citep[see, e.g.,][and references therein]{2005MNRAS.364...69S}.  \citet{2005MNRAS.364...69S} find that their most successful model, which best reproduces the current geometry and morphology of the two galaxies, has the following starting conditions.  The two galaxies are very roughly coplanar as is the grazing collision trajectory between them.  The disk planes of these galaxies are inclined only a modest amount (e.g., $\sim$$25^\circ-50^\circ$) with respect to the plane of the sky.  The best model results are obtained if the two galaxies had their closest approach during their first pass when IC 2163 was on the western side of NGC 2207  (see Fig.\,\ref{fig:montage}) some 300 Myr in the past. The orbit of the two galaxies is counterclockwise in Fig.\,\ref{fig:montage} as is the intrinsic rotation of IC 2163, i.e., the collision is prograde with respect to the more compact galaxy.  By contrast, the more expansive galaxy NGC 2207 is rotating the opposite way, and the collision is retrograde with respect to it.  The simulations show that the spiral arms of NGC 2207 were present before the collision, and are not much perturbed by the collision. Whereas, the prominent ``ocular'' feature in IC 2163 (see especially the {\em Spitzer} images) was created by the encounter, and its existence apparently shows that the collision can not have been going on for more than an orbit of the two galaxies.  Any star formation that the grazing collision has induced has likely been ongoing for the past few hundred Myr.  The total masses (including the dark-matter halos) of the two galaxies were taken to be $1.5 \times 10^{11}$ and $1.1 \times 10^{11}$ $M_\odot$ for NGC 2207 and IC 2163, respectively.

\begin{figure*}
\begin{center}
\includegraphics[width=0.8\linewidth]{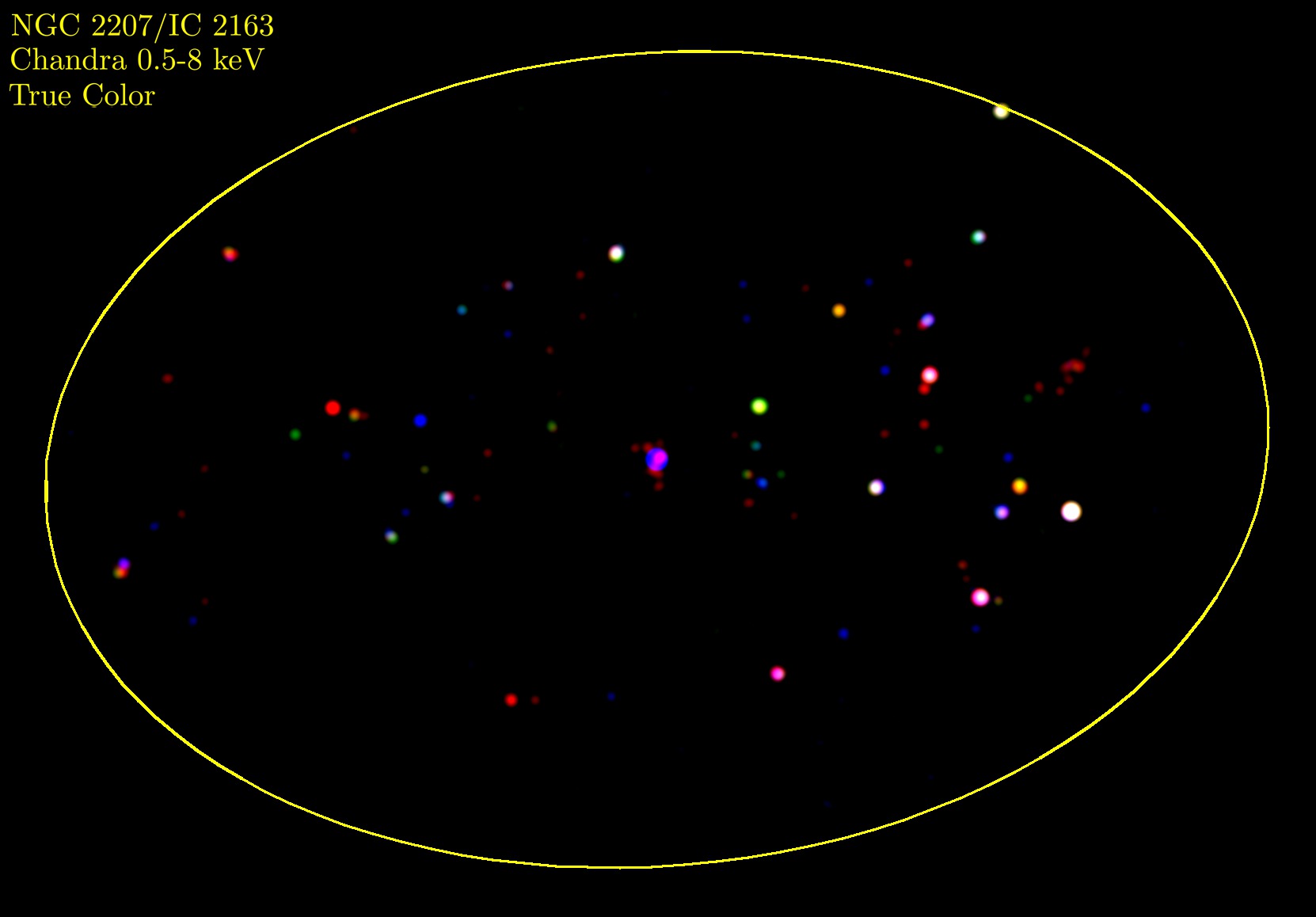} 
\caption{\textit{Chandra} X-ray image, smoothed with a Gaussian kernel of $\sigma = 2.5''$  of the galaxy pair NGC~2207/IC~2163. The red color corresponds to the soft (0.2 to 1.5 keV) band, the green color to the medium (1.5 to 2.5 keV) band and the blue color to the hard (2.5 to 8.0 keV) band. Note that the definition of ``soft" and ``hard" in the production this figure is different than that used in the data analysis and calculation of hardness ratio.  The D25 ellipse is superposed for reference. }
\label{fig:xray}
\end{center}
\end{figure*}

\section{X-ray analysis}
\label{sec:Xray_analysis}

\subsection{Data preparation}
We analyzed the publicly available \textit{Chandra}  ACIS-S observation of the galaxy pair NGC~2207 -- IC~2163, having identification number 11228 and an exposure time of $13$ ks. 
The data preparation was done following the standard CIAO\footnote{http://cxc.harvard.edu/ciao4.3/index.html} threads (CIAO version 4.3, CALDB version 4.5.1) for soft (0.5--2 keV), hard (2--8 keV) and broad (0.5--8.0 keV) energy bands. The point source detection was performed in the broad band using CIAO \texttt{wavdetect}, over the area within the $D25$ ellipse of NGC~2207 \& IC~2163. We used the $\sqrt{2}$-series from 1.0 to 8.0 as the scale parameter, in order to account for the variation of the effective width of the {\it Chandra} point spread function (PSF) from the inner to the outer parts of the analyzed observation, with reference to the {\em Chandra} aim point. We set the value of the parameter \texttt{sighthresh} as the inverse of the total number of pixels in the image ($\sim 10^{-6}$), in order to have one spurious detection per field. We used \texttt{maxiter} $= 10$, \texttt{iterstop} $= 0.00001$ and \texttt{bkgsigthresh} $= 0.0001$. The encircled fraction of source energy used for source estimation (parameter \texttt{eenergy}) was set to $0.8$.

We computed a monochromatic exposure map for the mean photon energy, i.e., 1.25 keV (soft band), 5.0 keV (hard band), and 4.25 keV (broad band), of each band. This was done using the CIAO script \texttt{fluximage}, which runs the tool \texttt{mkinstmap} to calculate the instrument map, i.e., the effective-area-weighted exposure map in instrument coordinates, for the center of each energy band, and \texttt{mkexpmap} to calculate the exposure maps in sky coordinates for each energy band. The exposure maps allow us to measure fluxes and hardness ratios for the detected X-ray point-like sources, using the photometric procedure described below.

\subsection{Source counts}
\label{sec:photometry}
For each detected point source we measured the count rate inside a circular region centered at the source central coordinates produced by \texttt{wavdetect}. The size of the circular region for individual sources was determined by requiring the encircled PSF energy to be 85\% of the total PSF energy. In order to do so, the PSF at the position of each point source was first constructed and then mapped into the World Coordinate System (WCS) reference frame of the relative point source image using the CIAO tasks \texttt{mkpsf} and \texttt{reproject\_image} respectively. The background regions were defined as annuli having an inner radius equal to the radius of the source region and an outer radius that is 3 times larger.
The corrected source counts and errors were obtained by performing aperture-corrected photometry, following the same method as in \citet{2007A&A...468...49V} and \citet{2012MNRAS.419.2095M}: 
\begin{equation}
\label{eq:src_counts}
S=\frac{(b-d)C-dQ}{b\alpha - d\beta}
\end{equation}
\begin{equation}
\label{eq:src_counts_err}
\sigma^{2}_S=\frac{(b-d)^{2}\sigma^{2}_C+d^{2}\sigma^{2}_Q}{(b\alpha - d\beta)^{2}}.
\end{equation}
Here $S$ is the number of net counts from the source, $C$ is the number of counts inside the source region and $Q$ is the number of counts in the background region, $\alpha$ is the integral of the PSF over the source region (expressed as a fraction of the total PSF energy), $\beta$ is the corresponding integral of the PSF over the source and background regions, $b$ is the integral of the (effective-area weighted) exposure over the source and background regions, and $d$ is the exposure integrated over the source region.

We found four compact sources having background regions that overlap their neighboring sources. In these cases the source count estimation is compromised and it was corrected as follows. For the overlapping point sources we defined the radius of the circular region that included 90\% of the encircled PSF. We excluded these regions from both the image and exposure map in order to respectively subtract the source contribution from the background counts and correct the source area. Finally, we again performed the aperture-corrected photometry described above using the corrected image and exposure map.

\subsection{Luminosities and hardness ratios}
\label{sec:lums}
X-ray fluxes in the 0.5--8.0 keV band were estimated from the count rates measured for each source as described in the previous section. A counts-to-erg conversion factor was obtained by first extracting the {\em combined} spectrum of all point sources detected within the $D25$ ellipse, except the central source in NGC~2207, which may be an active galactic nucleus (AGN) partially covered by dense clouds \citep{2006ApJ...642..158E, 2012AJ....144..156K}. A background spectrum was extracted from large regions located far from detected point sources but on the same CCD chip as the galaxy pair under study. Some of the background regions were located between point sources within the $D25$ ellipse in order to account for the diffuse emission contribution of the galaxy itself. 

Source and background spectra were created using the CIAO task \texttt{dmextract} and the associated weighted ARF and RMF files were made using, respectively, the tasks \texttt{mkwarf} and \texttt{mkrmf}. The average source spectrum was binned so as to have a minimum of 20 total counts (i.e., not background subtracted) per channel and thereby facilitate minimum-$\chi^2$ fitting. The background-subtracted spectrum was modeled as an absorbed power law using XSPEC v. 12.7.1b. The best-fit model was obtained for $n_{H}=2.88^{+0.15}_{-0.13}\times 10^{21}\,\rm{cm}^{-2}$ and photon power-law index $\Gamma = 2.08^{+0.34}_{-0.31}$ with $\chi^{2}=15.37$ for 21 degrees of freedom. The best-fit power-law index is consistent with the centroid of the power law photon index distribution for luminous X-ray compact sources in star-forming galaxies, $\Gamma=1.97\pm 0.11$ \citep{2004ApJS..154..519S}. The best-fit column density is a factor of $\sim$3 larger than the average Galactic $n_{H}$ in this direction seen in the Leiden/Argentine/Bonn (LAB) Survey of Galactic HI \citep{2005A&A...440..775K}.

The best-fit spectral model was used to convert the count rates of each of the detected point sources into fluxes ($\rm{erg}\,\rm{cm}^{-2}\,\rm{s}^{-1}$). X-ray luminosities were calculated assuming the distance of $39.6$ Mpc \citep{1982ApJ...254....1A}.

Hardness ratios were used to investigate the spectral properties of the sources detected in the $0.5-8$ keV band. The procedure described in Sect. \ref{sec:photometry} was applied to the reference source list in both soft ($S$: 0.5--2 keV) and hard ($H$: 2--8 keV) bands. The respective source counts were used to calculate the X-ray hardness ratio as:
\begin{equation}
\label{eq:hr}
\rm{HR} = \frac{\rm{H}-\rm{S}}{\rm{H}+\rm{S}}
\end{equation}

\begin{deluxetable*}{cccccccccc}
\tablewidth{0pt}
\tabletypesize{\scriptsize}
\tablecaption{\label{tab:xray} NGC2207/IC2163 X-Ray Source Properties}
\tablehead{
	\colhead{Source} &
	\colhead{$\alpha_{J2000}$} &
	\colhead{$\delta_{J2000}$}  & 
	\colhead{$0.5-8\,\rm{keV}$} &
	\colhead{Signif} &
	\colhead{$0.5-2\,\rm{keV}$} &
	\colhead{$2-8\,\rm{keV}$} &
	\colhead{HR} &
    	\colhead{$L_{\rm{X}}$} &
	\colhead{$F_{\rm{X}}$} \\
	\colhead{} & 
	\colhead{(deg)} &
	\colhead{(deg)} & 
      	\colhead{(cts)} &
      	\colhead{($\sigma$)} &
	\colhead{(cts)} &
	\colhead{(cts)} &
         \colhead{(cts)} &
	\colhead{($10^{38}\,\rm{erg}\,\rm{s}^{-1}$)} &
	\colhead{($10^{-14}\,\rm{erg}\,\rm{cm}^{-2}\,\rm{s}^{-1}$)}\\
	(1)  & (2) & (3) & (4) & (5) & (6) & (7) & (8)  & (9) & (10)
}
\startdata 
1 & 94.08436 & -21.3851 & $20 \pm 6.1$ & 8 & $14 \pm 5.4$ & $5.9 \pm 4$ & $-0.41 \pm 0.32$ & $24 \pm 7.3$ & $1.26 \pm 0.39$ \\
2 & 94.07186 & -21.3806 & $40 \pm 8.5$ & 18 & $29 \pm 7.5$ & $11 \pm 5.2$ & $-0.46 \pm 0.21$ & $47 \pm 10$ & $2.51 \pm 0.54$ \\
3 & 94.07053 & -21.3758 & $18 \pm 6.3$ & 9.4 & $7 \pm 4.6$ & $11 \pm 5.2$ & $0.23 \pm 0.38$ & $22 \pm 7.8$ & $1.18 \pm 0.41$ \\
4 & 94.06623 & -21.3757 & $91 \pm 12$ & 40 & $61 \pm 10$ & $30 \pm 7.5$ & $-0.35 \pm 0.13$ & $110 \pm 14$ & $5.73 \pm 0.76$ \\
5 & 94.07828 & -21.3743 & $27 \pm 7.4$ & 12 & $18 \pm 6.3$ & $8.8 \pm 4.9$ & $-0.35 \pm 0.29$ & $32 \pm 8.7$ & $1.68 \pm 0.46$ \\
6 & 94.06943 & -21.3743 & $20 \pm 6.1$ & 9.5 & $18 \pm 5.8$ & $2.1 \pm 3$ & $-0.80 \pm 0.27$ & $25 \pm 7.5$ & $1.34 \pm 0.40$ \\
$7\dagger$  & 94.09183 & -21.3727 & $76 \pm 11$ & 28 & $9.2 \pm 5.2$ & $66 \pm 10$ & $0.76 \pm 0.12$ & $90 \pm 13$ & $4.78 \pm 0.69$ \\
8 & 94.07010 & -21.3726 & $7.3 \pm 4.4$ & 3 & $3.7 \pm 3.6$ & $3.7 \pm 3.6$ & $0.00 \pm 0.69$ & $9.1 \pm 5.4$ & $0.48 \pm 0.29$ \\
9 & 94.11187 & -21.3697 & $14 \pm 5.4$ & 5.5 & $13 \pm 5.3$ & $1.2 \pm 2.8$ & $-0.83 \pm 0.37$ & $21 \pm 7.9$ & $1.10 \pm 0.42$ \\
10 & 94.08552 & -21.3696 & $27 \pm 6.8$ & 11 & $21 \pm 6.1$ & $6.7 \pm 4.1$ & $-0.51 \pm 0.25$ & $32 \pm 8.1$ & $1.72 \pm 0.43$ \\
11 & 94.07498 & -21.3679 & $33 \pm 7.4$ & 15 & $27 \pm 6.8$ & $6.6 \pm 4.1$ & $-0.60 \pm 0.21$ & $41 \pm 9.1$ & $2.19 \pm 0.48$ \\
12 & 94.10386 & -21.3641 & $7 \pm 4.3$ & 2.8 & $4.6 \pm 3.8$ & $2.4 \pm 3.2$ & $-0.32 \pm 0.70$ & $8.3 \pm 5.1$ & $0.44 \pm 0.27$ \\
13 & 94.08058 & -21.3642 & $14 \pm 5.2$ & 5.6 & $11 \pm 4.9$ & $2.3 \pm 3$ & $-0.67 \pm 0.39$ & $16 \pm 6.1$ & $0.86 \pm 0.33$ \\
14 & 94.09434 & -21.3608 & $26 \pm 6.8$ & 12 & $16 \pm 5.7$ & $9.5 \pm 4.7$ & $-0.27 \pm 0.28$ & $31 \pm 8.1$ & $1.64 \pm 0.43$ \\
15 & 94.07195 & -21.3599 & $16 \pm 5.5$ & 7.8 & $9.9 \pm 4.6$ & $6.5 \pm 4$ & $-0.20 \pm 0.37$ & $19 \pm 6.6$ & $1.03 \pm 0.35$ \\
16 & 94.07056 & -21.3527 & $29 \pm 6.9$ & 14 & $21 \pm 6.1$ & $7.5 \pm 4.3$ & $-0.48 \pm 0.24$ & $35 \pm 8.3$ & $1.84 \pm 0.44$ \\
17 & 94.10830 & -21.3771 & $15 \pm 5.6$ & 5.6 & $7.1 \pm 4.3$ & $8.1 \pm 4.5$ & $0.07 \pm 0.41$ & $21 \pm 7.6$ & $1.10 \pm 0.40$ \\
18 & 94.10480 & -21.3749 & $17 \pm 6.1$ & 6.2 & $9.3 \pm 4.8$ & $8.1 \pm 4.6$ & $-0.07 \pm 0.39$ & $21 \pm 7.4$ & $1.12 \pm 0.39$ \\
19 & 94.07529 & -21.3686 & $7.5 \pm 4.3$ & 3.4 & $6.5 \pm 4.1$ & $0.94 \pm 2.6$ & $-0.75 \pm 0.63$ & $9.2 \pm 5.3$ & $0.49 \pm 0.28$ \\
20 & 94.07520 & -21.3648 & $9.9 \pm 5.3$ & 8.5 & $4.1 \pm 4.1$ & $5.7 \pm 4.4$ & $0.16 \pm 0.61$ & $12 \pm 6.5$ & $0.65 \pm 0.35$ \\
21 & 94.11826 & -21.3609 & $17 \pm 5.8$ & 7.4 & $13 \pm 5.2$ & $4.6 \pm 3.7$ & $-0.47 \pm 0.35$ & $24 \pm 8$ & $1.28 \pm 0.42$ \\
22 & 94.10091 & -21.3865 & $8.3 \pm 4.5$ & 3.3 & $8.3 \pm 4.5$ & $0 \pm 2.2$ & $-1.00 \pm 0.53$ & $12 \pm 6.6$ & $0.65 \pm 0.35$ \\
23 & 94.08575 & -21.3719 & $8 \pm 4.3$ & 3 & $4.5 \pm 3.6$ & $3.4 \pm 3.4$ & $-0.14 \pm 0.62$ & $9.4 \pm 5.1$ & $0.50 \pm 0.27$ \\
24 & 94.11424 & -21.3712 & $6.2 \pm 4.2$ & 2.3 & $5 \pm 4$ & $1.2 \pm 2.9$ & $-0.60 \pm 0.79$ & $9 \pm 6.1$ & $0.48 \pm 0.33$ \\
25 & 94.11039 & -21.3701 & $11 \pm 5.1$ & 5 & $10 \pm 5$ & $1 \pm 2.8$ & $-0.81 \pm 0.46$ & $14 \pm 6.5$ & $0.76 \pm 0.35$ \\
26 & 94.10103 & -21.3627 & $10 \pm 4.9$ & 4.6 & $5.5 \pm 4$ & $4.5 \pm 3.8$ & $-0.10 \pm 0.55$ & $12 \pm 5.8$ & $0.63 \pm 0.31$ \\
27 & 94.12492 & -21.3790 & $30 \pm 7.3$ & 9.9 & $22 \pm 6.5$ & $7.2 \pm 4.3$ & $-0.51 \pm 0.25$ & $37 \pm 9$ & $1.95 \pm 0.48$ \\
$28\ddagger $ & 94.06592 & -21.3673 & $6.8 \pm 4.8$ & 5.8 & $7.3 \pm 4.8$ & $0 \pm 2.3$ & $-1.00 \pm 0.64$ & $8 \pm 5.7$ & $0.43 \pm 0.30$ \\
\enddata
\tablecomments{(1) Source number, (2) Right Ascension (RA), (3)
  Declination (Dec). (4) Net counts in broad (0.5--8 keV) band,
  computed with eqns. (\ref{eq:src_counts}) and
  (\ref{eq:src_counts_err}). The uncertainty expressed here takes into
  account the fluctuations in the source as well as in the
  background. (5) Broad band source detection significance from {\tt
    wavdetect}.  This computes how unlikely it is for the background
  in the customized psf region to fluctuate to yield the detected
  number of counts.  Note that the psf region is optimized differently
  in {\tt wavdetect} than in the calculation of column (4) and is
  typically larger than in the latter. (6)-(7) Net counts in soft
  (0.5--2 keV) and hard (2--8 keV) bands respectively, computed with
  eq. (\ref{eq:src_counts}). Uncertainties in net counts are quoted to
  $1\,\sigma$ and were obtained by mean of
  eq. (\ref{eq:src_counts_err}). (7) Hardness ratio, computed with eq. (\ref{eq:hr}). Uncertainties were obtained by applying error propagation to the uncertainties in the net counts. (8) X-ray luminosity in the 0.5--8 keV band, (9) X-ray flux in the 0.5--8 keV band. $\dagger$ Central Active Galactic Nucleus, $\ddagger$ Extended soft X-ray source.}
\end{deluxetable*}

\section{Luminous X-Ray Sources}

\subsection{Discrete Source Content}

Following the method of  \citet{2006A&A...447...71V}, we computed the completeness function $K(L_{\rm{X}})$ of the present \textit{Chandra} observations of the $D25$ region. At the assumed distance of $39.6$ Mpc, the luminosity, for which $> 80\%$ of point sources are detected in the 0.5--8 keV band, is $L_{\rm{comp}} \simeq 1.0 \times 10^{39} \rm{erg}\,\rm{s}^{-1}$ (equivalently, $K(L_{\rm{comp}}) = 0.8$).

The properties of the discrete luminous X-ray sources that our analysis yielded for NGC~2207/~IC 2163 are summarized in Table \ref{tab:xray}. In particular, for all sources detected within the $D25$ ellipse, we list the source location, the net counts after background subtraction in several bands, the hardness ratio (defined above), the X-ray luminosity, and the X-ray flux. 
The source counts in the 0.5--8 keV band were computed from eqns. (1) and (2), and their associated uncertainties are listed in column (4).  These values are used to compute the corresponding source fluxes and luminosities, as well as their uncertainties (columns (10) and (9)).  By contrast, the detection significances listed in column (5) were calculated by {\tt wavdetect}.  Most of the detection significances have a sensible correspondence with the source-count uncertainties listed in column (4); however, the former expresses the probability that the background could fluctuate to yield the number of observed counts, whereas the latter also includes the fluctuations in the source counts themselves (for brighter sources this is actually the dominant contribution).  The two columns are in reasonable accord given the different questions that they address.  A possible exception is source no. 28.  According to the {\tt wavdetect} output, the net source counts for this object in the 0.5--8 keV band are $13.9 \pm 3.9$, while the detection significance is $5.8\sigma$.  This is attributed to the fact that source region defined by {\tt wavdetect} is larger (by nearly a factor of 2) than what we used to derive the results in column (4) from eqns. (1) and (2).  We note, however, that we use the {\tt wavdetect} output only to ascertain the existence of a source and to determine the source coordinates.  By contrast, we perform X-ray photometry, as described in Section~\ref{sec:photometry}, utilizing somewhat different source regions than those defined by {\tt wavdetect}; therefore, the detection significances in column (5) do not exactly correspond to the net source counts listed in column (4). We select the source sample to be analyzed by means of the incompleteness analysis, which accounts for the sensitivity variations across the image.

The {\em Chandra} X-ray image is shown in Fig.\,\ref{fig:xray}.

In all, 28 sources were detected, one of which (source no.~7) is likely a low-luminosity AGN associated with NGC 2207 \citep{2012AJ....144..156K}, and 6 are just below our completeness threshold (as well as below the `ULX limit' of $10^{39}$ erg s$^{-1}$; see \S3.2). A total of 21 sources are sufficiently bright to be ULXs. We note that source no. 28 has a soft spectrum, and may be part of an elongated soft X-ray feature. It is located in the outer spiral arm of NGC~2207, $\sim$$1.5\arcmin$ N-W from its center at the location of the dusty starburst region called {\em feature i} \citep{2000AJ....120..630E}.

\begin{figure*}
\begin{center}
\includegraphics[width=1.0\linewidth]{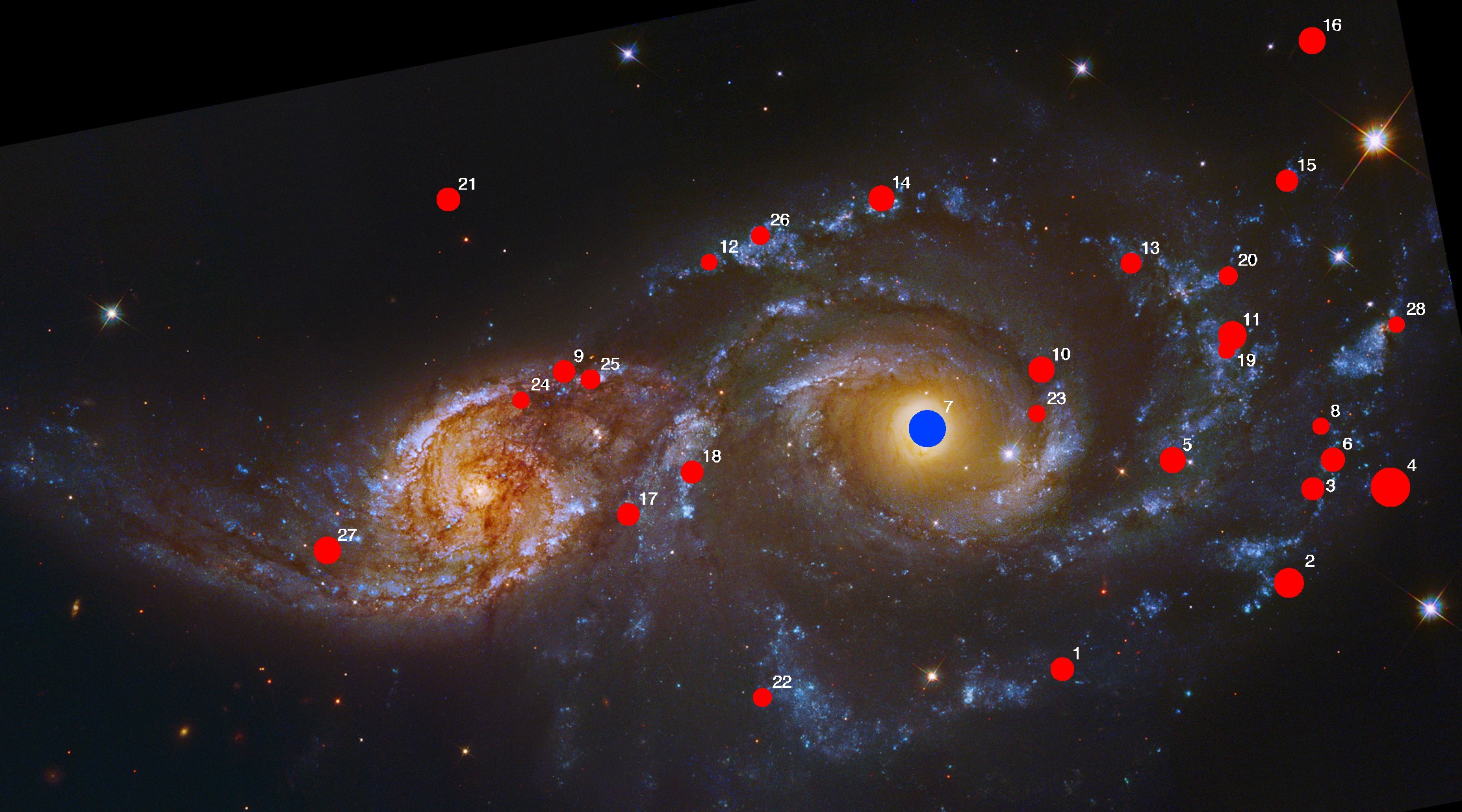}
\caption{The HST image of NGC~2207/IC~2163 with the filled red circles superposed indicating the locations of the X-ray point sources detected above the completeness limit. The size of each circle is proportional to the cube root of the luminosity of the individual X-ray source.  Each circle is annotated according to the source number in Table 1.}
\label{fig:hst_src}
\end{center}
\end{figure*}

In order to show the distribution of the ULX population with respect to the morphological structures of the galaxy pair, we plot the X-ray point sources detected above the completeness limit, superposed on the HST image of NGC2207/IC2163 in Figure \ref{fig:hst_src}. The red circles indicate the location of individual ULXs, the circle size being proportional to the cube root of the 0.5--8 keV luminosity of the given X-ray source.

\subsection{X-ray luminosity function}
\label{sec:xlf}
\begin{figure}
\includegraphics[width=1.0\columnwidth]{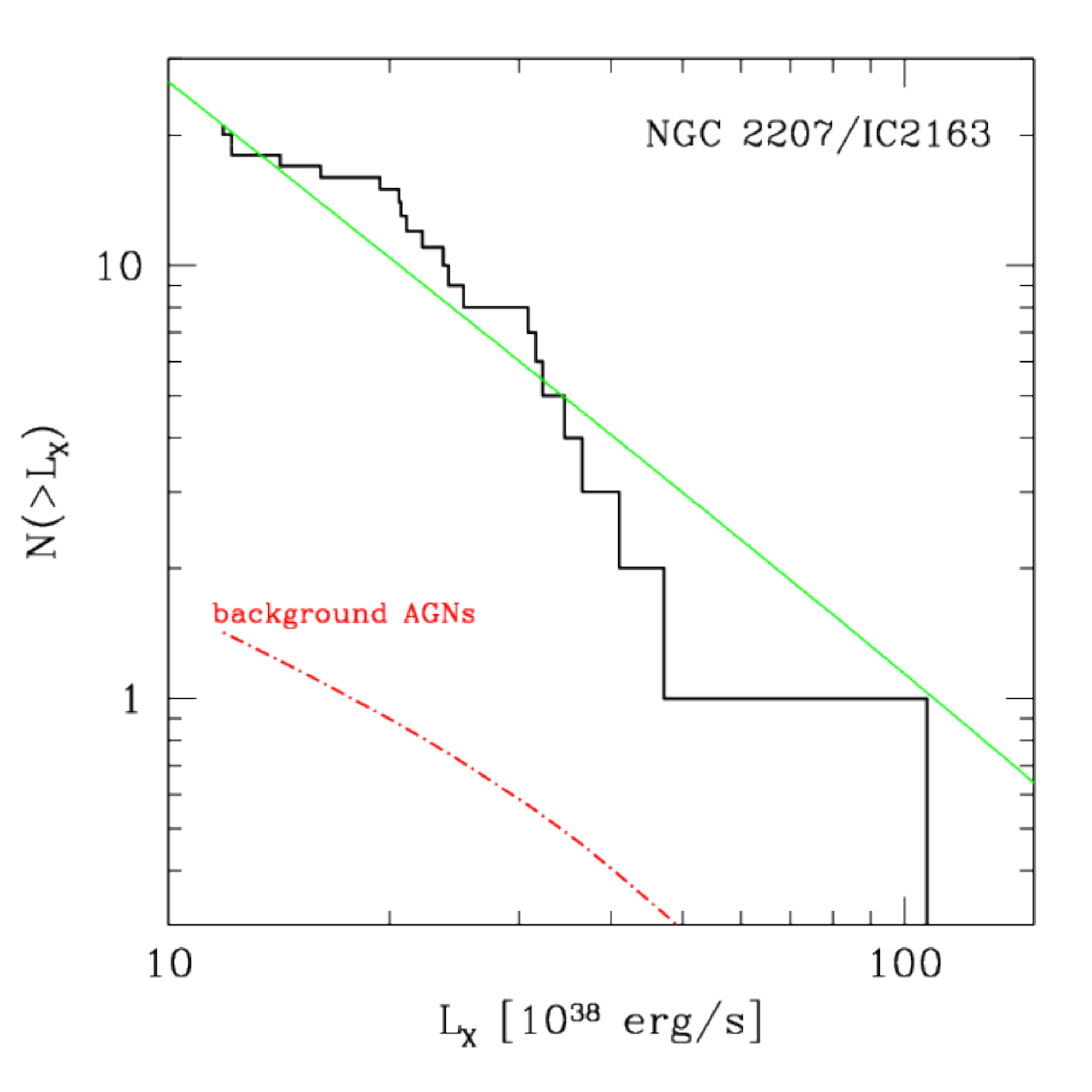}
\caption{Cumulative X-ray luminosity distribution of the compact sources detected within the $D25$ ellipse, above the completeness luminosity of the galaxy pair NGC~2207/IC~2163 (solid, black line). The central AGN is not included. The dash-dotted (red) curve shows the predicted level of resolved background AGNs, based on results from \citet{2008MNRAS.388.1205G}. The green solid line is the best fitting power law model.}
\label{fig:xlf}
\end{figure}

Twenty one compact sources with luminosities above $L_{\rm{comp}}$ were detected within the $D25$ region. Their luminosities range from $1.2\times 10^{39}\,\rm{erg}\,\rm{s}^{-1}$ to $1.1\times 10^{40}\,\rm{erg}\,\rm{s}^{-1}$. We constructed the cumulative X-ray luminosity function (XLF), and it is shown in Figure \ref{fig:xlf}. We modeled the XLF with a single-slope power law with a high luminosity cut-off exceeding the luminosity of the brightest compact source detected in the galaxy pair:
\begin{equation}
\frac{dN}{dL_{38}}=\xi \times L_{38}^{-\gamma}, ~~~~~L\le 10^{41}\,\rm{erg}\,\rm{s}^{-1}.
\label{eq:xlf}
\end{equation} 
A fit of the cumulative XLF using a maximum likelihood method yielded a slope of $\gamma = 2.35^{+0.33}_{-0.29}$. We performed a Kolmogorov-Smirnov (KS) test to determine the goodness of fit. The $D$ value obtained for the KS test is 0.24 for 21 sources corresponding to a significance level of $\sim$15\% that the data and model are from different distributions. This indicates that the model describes the data fairly well.

This slope is steeper than the slope of $\sim$$1.6$ that is typically found for high-mass X-ray binary (HMXB) luminosity distributions below $\sim$$10^{40}\,\rm{erg}\,\rm{s}^{-1}$ \citep{2003MNRAS.339..793G,2011ApJ...741...49S, 2012MNRAS.419.2095M}. On the other hand, due to the limited sensitivity of this {\it Chandra} observation we may be sampling only the roll-off of the power-law distribution that extends beyond the above mentioned slope at lower luminosities to values of $L_{\rm{X}}$ in the range of $\sim$$1-3 \times10^{39}$ erg s$^{-1}$. A similar XLF slope, of $\gamma=2.62^{+0.64}_{-0.50}$, is observed in one of the star-forming galaxies in the sample of \citet{2012MNRAS.419.2095M}, NGC~3079 for compact sources detected above a completeness limit of $\sim$$10^{38}\,\rm{erg}\,\rm{s}^{-1}$.

Interestingly, there is a lack of sources brighter than $\sim$$10^{40}\,\rm{erg}\,\rm{s}^{-1}$ in NGC 2207/IC 2163. On the other hand, assuming the HMXB luminosity function from \citet{2012MNRAS.419.2095M}, and rescaling it to match the observed number of sources in NGC~2207/IC~2163, the predicted number of HMXBs above $10^{40}\,\rm{erg}\,\rm{s}^{-1}$ is $2.4 \pm 1.5$ (assuming a Poisson distribution), which is fairly consistent with what we observe. Luangtip et al. (in preparation) found the same evidence of  a lack of sources brighter than $\sim$$10^{40}\,\rm{erg}\,\rm{s}^{-1}$ in a sample of 17 luminous infrared galaxies (LIRGs) located at distances between $\sim$$14$ and $\sim$$61$ Mpc. This may be due to the fact that the brightest ULXs should have very short lifetimes (e.g., $\lesssim$ few $\times 10^6$ yr; see Madhusudhan et al.~2008), and therefore their numbers and the concomitant position of the break in the XLF may be dependent on the very recent SFR history of the host galaxy.

We predicted the contribution of background AGNs within the $D25$ ellipse above the completeness luminosity based on the work of \citet{2008MNRAS.388.1205G}. Their $\log N-\log S$ function for the 0.5--10 keV band was converted to apply in the broad band of 0.5--8 keV. The result is that $\sim$$1.7$ background AGNs with $L_{\rm{X}} > L_{\rm{comp}}$ are expected to be present and to have a combined luminosity $L_{\rm{AGN}}(>L_{\rm{comp}}) = 7.32\times 10^{39}\,\rm{erg}\,\rm{s}^{-1}$. In this work AGN ``luminosities" are computed as $4\pi\,\rm{D}^{2}\,S_{\rm{AGN}}(>S_{\rm{comp}})$, where $D$ is the distance to the galaxy pair and $S_{\rm{AGN}}$ is the predicted total flux of background AGNs above the completeness threshold flux $S_{\rm{comp}}$. The cumulative luminosity distribution of background AGNs is marked in Fig.~\ref{fig:xlf} by a dot-dashed (red) curve.  Thus, perhaps one or two of the 22 detected sources may actually be background AGNs. In order to plot the background AGN contribution to the luminosity function we convert their flux to "luminosity" as defined as above. Obviously, although this quantity has units of erg/s, it has nothing to do with the true luminosities of background objects. However, the introduction of this quantity simplifies the calculation of contributions of background objects to the numbers of sources and their total luminosity.

\section{Star formation rate}
\label{sec:sfr}

The SFR is one of the most important parameters in the investigation of gas-rich galaxies. Over the last decade, several studies have demonstrated the existence of a tight correlation between the {\em collective} number of luminous X-ray sources and the integrated SFR of late-type host galaxies \citep[e.g.,][]{2003MNRAS.339..793G, 2010MNRAS.408..234M, 2011ApJ...741...49S, 2012MNRAS.419.2095M}. This relation can now be explored in greater detail using spatially resolved images of SFR surface density. Such SFR images can be constructed from {\it Spitzer} and {\it Galex} archival data, following a recent technique introduced by \citet{2008AJ....136.2782L}. This will allow us to investigate the spatial distribution of the X-ray point sources and their luminosities as a function of the {\em local} SFR at their location.  

\subsection{Star formation rate surface density images}
\label{sec:sfr_map}

\begin{figure*}
\centering
\includegraphics[width=0.8\linewidth]{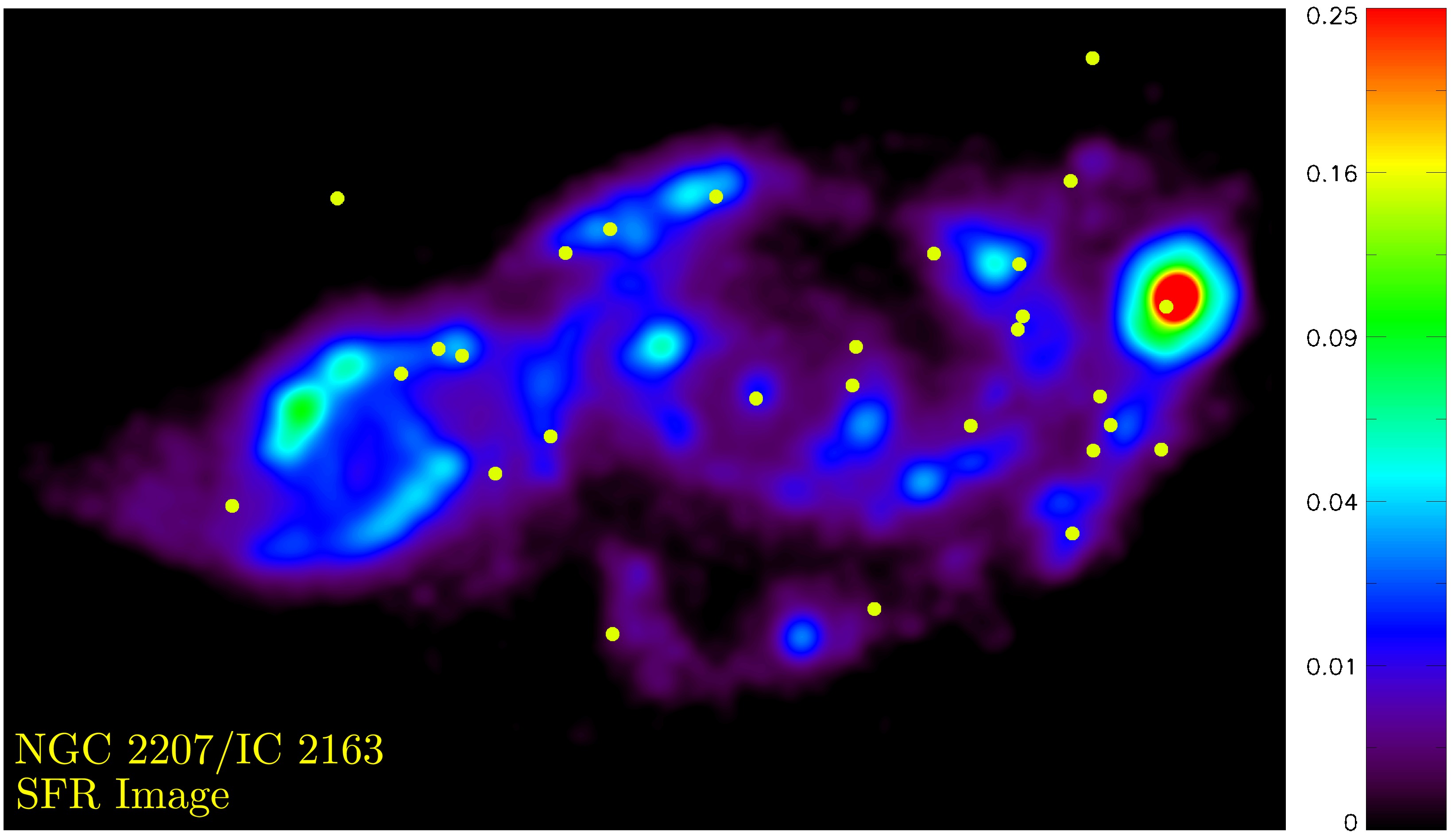}
\caption{SFR density map for NGC~2207/IC~2163, in units of $M_{\odot}\,\rm{yr}^{-1}\,\rm{kpc}^{-2}$, obtained by combining {\it Galex} FUV and {\it Spitzer} $24\,\mu\rm{m}$ images according the prescription of Leroy et al.~(2008). The small yellow circles mark the locations of the X-ray sources.  
See Sect. \ref{sec:sfr_map} for details.}
\label{fig:sfrmap}
\end{figure*}

We adopted the recipe provided by \citet{2008AJ....136.2782L} to estimate the spatially-resolved SFR distribution in the NGC2207/IC2163 system. The SFR density in units of $M_\odot~{\rm yr}^{-1}~{\rm kpc}^{-2}$, was therefore estimated using their eq. (D11):
\begin{equation}
\label{eq:sfr_density}
\Sigma_{\rm{SFR}} = 8.1\times 10^{-2} I_{FUV}+3.2\times 10^{-3} I_{24\,\mu m}
\end{equation}
where $I_{FUV}$ and $I_{24\,\mu m}$ are in units of $\rm{MJy}\,\rm{ster}^{-1}$. The combination of {\it Galex} FUV and {\it Spitzer} MIPS $24\,\mu\rm{m}$ images is relatively easy to implement because of their reasonably similar angular resolutions ($4\arcsec$ and $6\arcsec$ FWHM, respectively) and sensitivities. The {\it Spitzer} $24\,\mu\rm{m}$ images are already calibrated in $\rm{MJy}\,\rm{ster}^{-1}$. To convert the {\it Galex} FUV images from count rate (count $s^{-1}$) to $\rm{MJy}\,\rm{ster}^{-1}$ we used the conversion factor $C_{\rm{FUV}}=2.028$. This results from converting counts s$^{-1}$ to flux (erg cm$^{-2}$ s$^{-1}$ $\AA^{-1}$)\footnote{http://galexgi.gsfc.nasa.gov/docs/galex/FAQ/counts\_background.html}, using a factor of $1.40\times 10^{-15}$, and a successive conversion from erg cm$^{-2}$ s$^{-1}$ $\AA^{-1}$ to $\rm{MJy}\,\rm{ster}^{-1}$, involving a factor of $1.45\times 10^{15}$.

Importantly, this SFR estimator is sensitive to both dust-obscured and exposed star formation activity. As the FUV emission originates from the photospheres of O and B stars, {\em Galex} intensities tend to indicate the somewhat older regions of star formation ($\tau_{\rm{FUV}}\gtrsim 30$ Myr), where the obscuring dust has already been cleared. The $24\,\mu\rm{m}$ emission originates from dust grains heated by embedded young ionizing stars and traces the star formation over timescales, $\tau_{24\,\mu\rm{m}}\sim 5-10$ Myr \citep[][Rappaport et al.~2012, in preparation; Mark Krumholz, private communication 2012]{2007ApJ...666..870C}. The SFR over these shorter timescales may be more closely related to the upper end of the ULX luminosities (see Sect. \ref{sec:sfr_distr}).

For the first term of eq.~(\ref{eq:sfr_density}) we used publicly available {\it Galex} far-ultraviolet (FUV, 1529\,\AA) background-subtracted images from the All Sky Surveys (AIS) program\footnote{http://galex.stsci.edu/GR4/?page=mastform}. These images are calibrated in units of counts per pixel per second, and are also corrected for the relative instrumental response. The units of the FUV image were converted into $\rm{MJy}\,\rm{ster}^{-1}$ prior to combining the latter with the $24\,\mu\rm{m}$ image.

The second term in eq.~(\ref{eq:sfr_density}) was based on a {\em Spitzer} MIPS $24\,\mu\rm{m}$ Large Field image. We used the ``post Basic Calibrated Data" products which are calibrated in $\rm{MJy}\,\rm{ster}^{-1}$ and suitable for photometric measurements\footnote{http://irsa.ipac.caltech.edu/applications/Spitzer/Spitzer/}.  We measured the $24 \,\mu\rm{m}$ background in a region away from the galaxy, and subtracted it from the image before combining the latter with the FUV map. As the pixel scales of the $24\,\mu\rm{m}$ and FUV images are different ($2.45 \arcsec/\rm{pix}$ and $1.5 \arcsec/\rm{pix}$, respectively), we spatially interpolated the $24\,\mu\rm{m}$ image in order to match the better resolution of the {\it Galex} FUV image. This was done using the routine HASTROM, from the NASA IDL Astronomy User's Library\footnote{http://idlastro.gsfc.nasa.gov/}. The routine properly interpolates without adding significant spatial information (i.e., no spatial frequency content beyond the intrinsic resolution of the original image). In essence, the $24\,\mu\rm{m}$ image was simply oversampled. The resulting SFR surface density map is displayed in Figure \ref{fig:sfrmap} at the same spatial resolution and with the same pixel coordinates as the {\it Galex} FUV image (but limited by the {\em Spitzer} resolution).

Interestingly, the SFR map shows a similar morphology to the radio emission map at 4.86~GHz, dominated by nonthermal (synchrotron) radiation, discussed by \citet{2011A&A...533A..22D} (see their Figs. 1 and 2).

\subsection{Integrated star formation rate}
\label{sec:int_sfr}
We calculated the overall SFR of the galaxy pair NGC~2207/IC~2163 by integrating eq.~(\ref{eq:sfr_density}) inside the $D25$ region. The resulting equation corresponds to the fiducial method from \citet{2008AJ....136.2782L}, in particular to their eq.~D10, on which their prescription for obtaining star formation rate surface density maps (their eq.~D11), is based.

The FUV and $24\,\mu \rm{m}$ terms for the integrated SFR were obtained by \citet{2008AJ....136.2782L} assuming a stellar initial mass function (IMF) as in \citet{2007ApJ...666..870C} (i.e., slope $-1.3$ for the $0.1-0.5\,M_{\odot}$ mass range and $-2.3$ for $0.5-120\,M_{\odot}$).
Using this prescription, we obtained a total SFR of $11.8 \,M_{\odot}\,\rm{yr}^{-1}$ integrated within the $D25$ region of the galaxy pair (for the \cite{2007ApJ...666..870C} IMF). The corresponding value of SFR based on a Salpeter IMF from 0.1 to 100 $M_{\odot}$, would be $18.8 \,M_{\odot}\,\rm{yr}^{-1}$ (the IMF conversion yields a factor of 1.59 difference). For comparison, the integrated SFR of the star-forming galaxies NGC~4194, NGC~7541 and the Cartwheel \citep[see, e.g.,][]{2004A&A...426..787W,2009A&A...501..445C}, assuming a Salpeter IMF as above, is $16.8\,M_{\odot}\,\rm{yr}^{-1}$,  $14.7\,M_{\odot}\,\rm{yr}^{-1}$ and $17.6\,M_{\odot}\,\rm{yr}^{-1}$ respectively \citep{2012MNRAS.419.2095M}.

\section{Spatially-resolved $N_{\rm{X}}-\rm{SFR}$, $L_{\rm{X}}-\rm{SFR}$ relations} 
\label{sec:nx_lx_sfr_rel}

\begin{figure}
\begin{center}
\includegraphics[width=0.99\columnwidth]{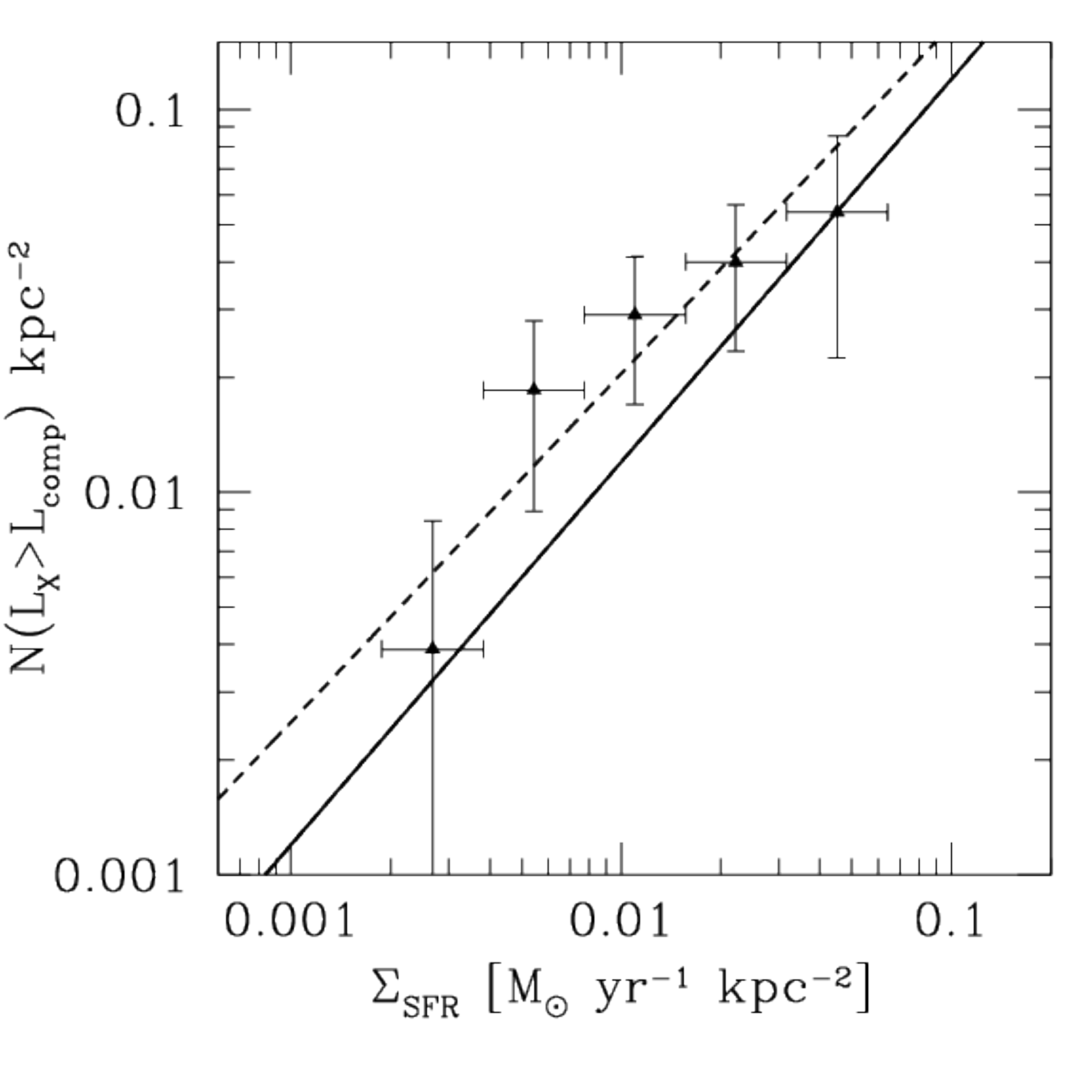} \vglue-0.1cm
\includegraphics[width=0.99\columnwidth]{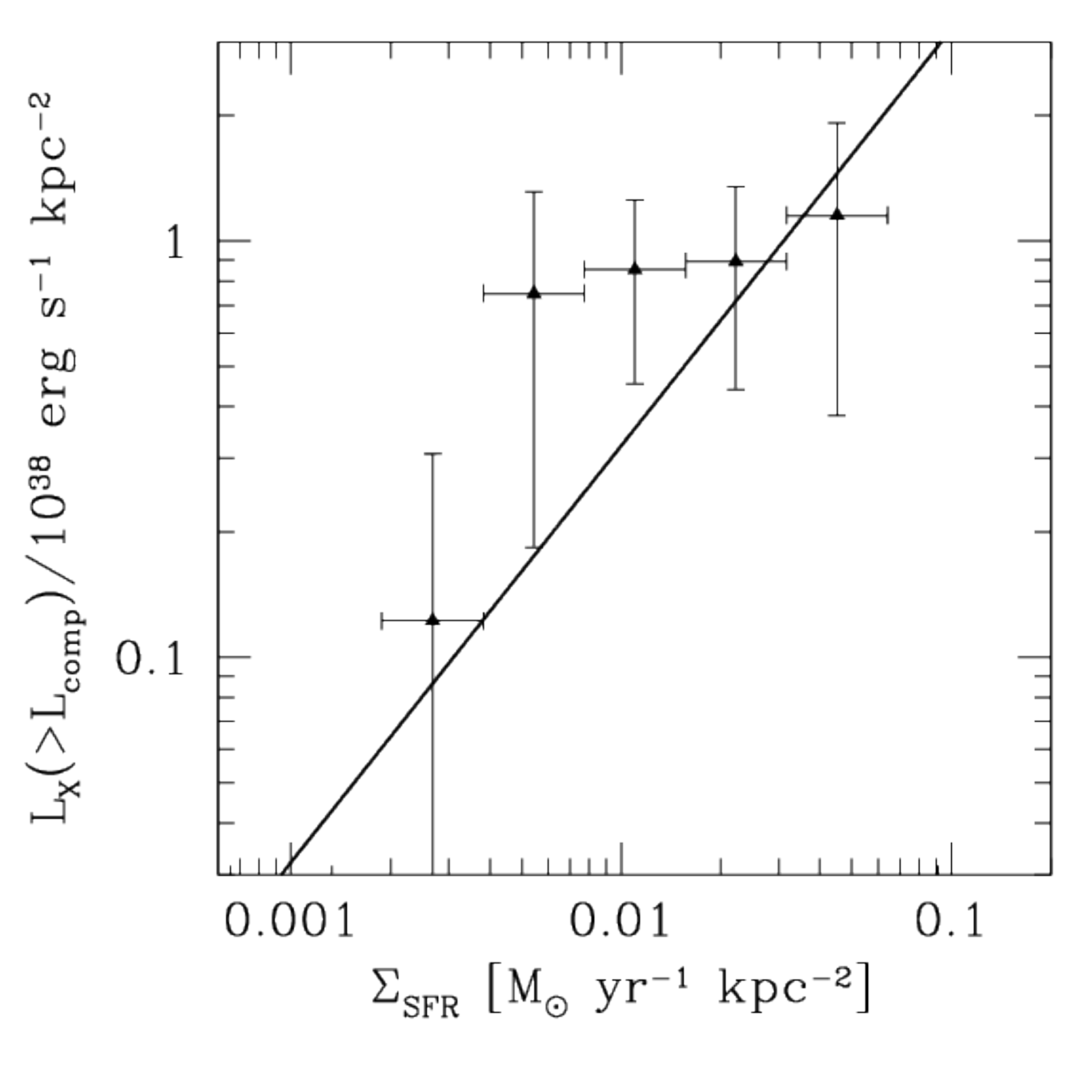} \vglue-0.1cm
\caption{Relation between the {\em local} SFR density in NGC~2207/IC~2163 ($M_{\odot}\,\rm{yr}^{-1}\,\rm{kpc}^{-2}$) and the number density ($N_{\rm{X}}/\rm{kpc}^{2}$, top panel) and luminosity density ($L_{\rm{X}}/\rm{kpc}^{2}$, lower panel) of luminous X-ray sources above our completeness threshold in $L_{\rm{X}}$. Both these quantities are corrected for the contribution of background AGNs. The star formation rate density was computed from the \citet{2008AJ....136.2782L} algorithm as described in Sect.\ref{sec:sfr_map}. The solid curves are the galaxy-wide average $N_{\rm{X}}-\rm{SFR}$ (top panel) and $L_{\rm{X}}-\rm{SFR}$ (bottom panel) relations for HMXBs obtained in \citet{2012MNRAS.419.2095M} (their eqs.(20) and (22) respectively). We show them for comparison, rescaled in order to match the IMF assumption in the SFR recipe from \citet{2008AJ....136.2782L}. For these expressions, $N_{\rm{X}}$ and $L_{\rm{X}}$ are $N,\,L (L_{\rm{X}}\gtrsim 10^{39}\,\rm{erg}\,\rm{s}^{-1}$), as is the case for the observational data. The dashed line in the top panel is the relation between the galaxy-wide average numbers of ULXs and SFR by \citet{2010MNRAS.408..234M} (their eq. (6)), which is slightly non-linear.}
\label{fig:nx_lx_sfr}
\end{center}
\end{figure}

We investigated the occurrence of the X-ray point sources and their luminosities as a function of the {\em local} SFR. We started from the SFR density image obtained as described in Sect.~\ref{sec:sfr_map}. A set of SFR density bins with constant logarithmic spacing was defined. The source number and their collective luminosity above $L_{\rm{comp}}$ (respectively $N_{\rm{X}}(L > L_{\rm{comp}})$ and $L_{\rm{X}}(> L_{\rm{comp}})$) were assigned to each bin of SFR density according to the SFR density value at the position of the source. We counted the number of pixels in each bin of SFR density and their cumulative area. Knowing the area, and based on the $\log{N}-\log{S}$ function from \citet{2008MNRAS.388.1205G}, we calculated the predicted number of background AGNs, $N_{\rm{AGN}}(L> L_{\rm{comp}})$, and their luminosity, $L_{\rm{AGN}}(>L_{\rm{comp}})$ above the completeness luminosity threshold, in accord with the procedure used in \S \ref{sec:xlf}. Typically, this amounts to less than one background source per bin. These two quantities were subtracted to yield $N_{\rm{X}}(L> L_{\rm{comp}})$ and $L_{\rm{X}}(> L_{\rm{comp}})$ respectively and the resulting value was divided by the total area in $\rm{kpc}^{2}$ in each SFR density bin. 

The final values of surface density of X-ray point sources (sources/kpc$^{2}$) and luminosity (erg s$^{-1}$/kpc$^{2}$) corrected for background AGNs, are plotted against the value of the SFR surface density in Fig.~\ref{fig:nx_lx_sfr}. Pixels with SFR density less than $6\times 10^{-4}\,M_{\odot}\,\rm{yr}^{-1}\,\rm{kpc}^{-2} $ were not used as they are dominated by background noise. The latter was measured in two large regions of the SFR density image outside the $D25$ ellipse. The resulting mean values in the two regions are $(6.4\pm 1.3)\times 10^{-5}$ and $(7.3\pm 0.9)\times 10^{-5}\,M_{\odot}\,\rm{yr}^{-1}\,\rm{kpc}^{-2}$ with $rms$ in the range of $6.4-6.8\times 10^{-4}$.

For comparison with the present measurements, in Fig.~\ref{fig:nx_lx_sfr} we also plot the multiple-galaxy-wide average $N_{\rm{X}}-\rm{SFR}$ and $L_{\rm{X}}-\rm{SFR}$ relations for HMXBs obtained in \citet{2012MNRAS.419.2095M} (their eqs.(20) and (22)). Here $N_{\rm{X}} = N(L_{\rm{X}}\gtrsim 10^{39}$ erg s$^{-1}$) as for the observed data. Since the SFR in the \citet{2012MNRAS.419.2095M} relation is based on the Salpeter IMF from 0.1 to 100 $M_{\odot}$, we first adjusted it to be consistent with the IMF assumed in the \citet{2008AJ....136.2782L} algorithm (see Sect. \ref{sec:int_sfr} for details). We also compare our results with the $N_{\rm{ULX}}-\rm{SFR}$ relation from \citet{2010MNRAS.408..234M}, whose definition of $N_{\rm{ULX}}$ is equal to our $N_{\rm{X}}$. The dashed line in the top panel of Fig.~\ref{fig:nx_lx_sfr} shows their eq.~(6), which is slightly non-linear. 

Fig.~\ref{fig:nx_lx_sfr} shows that the global relation between cumulative number of X-ray point sources and the integrated SFR of the host galaxy also holds on {\em local} scales (top panel). The small-number statistics involved did not allow us to study in any detail the $L_{\rm{X}}-\rm{SFR}$ relation (bottom panel). A more extensive work by Mineo et al. (in preparation), based on a significantly larger number of X-ray sources detected in a sample of nearby grand-design spiral galaxies, may yield more useful information on the linearity of the latter relation.

\section{SFR density distribution around ULXs}
\label{sec:sfr_distr}

\begin{figure}
\begin{center}
\includegraphics[width=1.0\columnwidth]{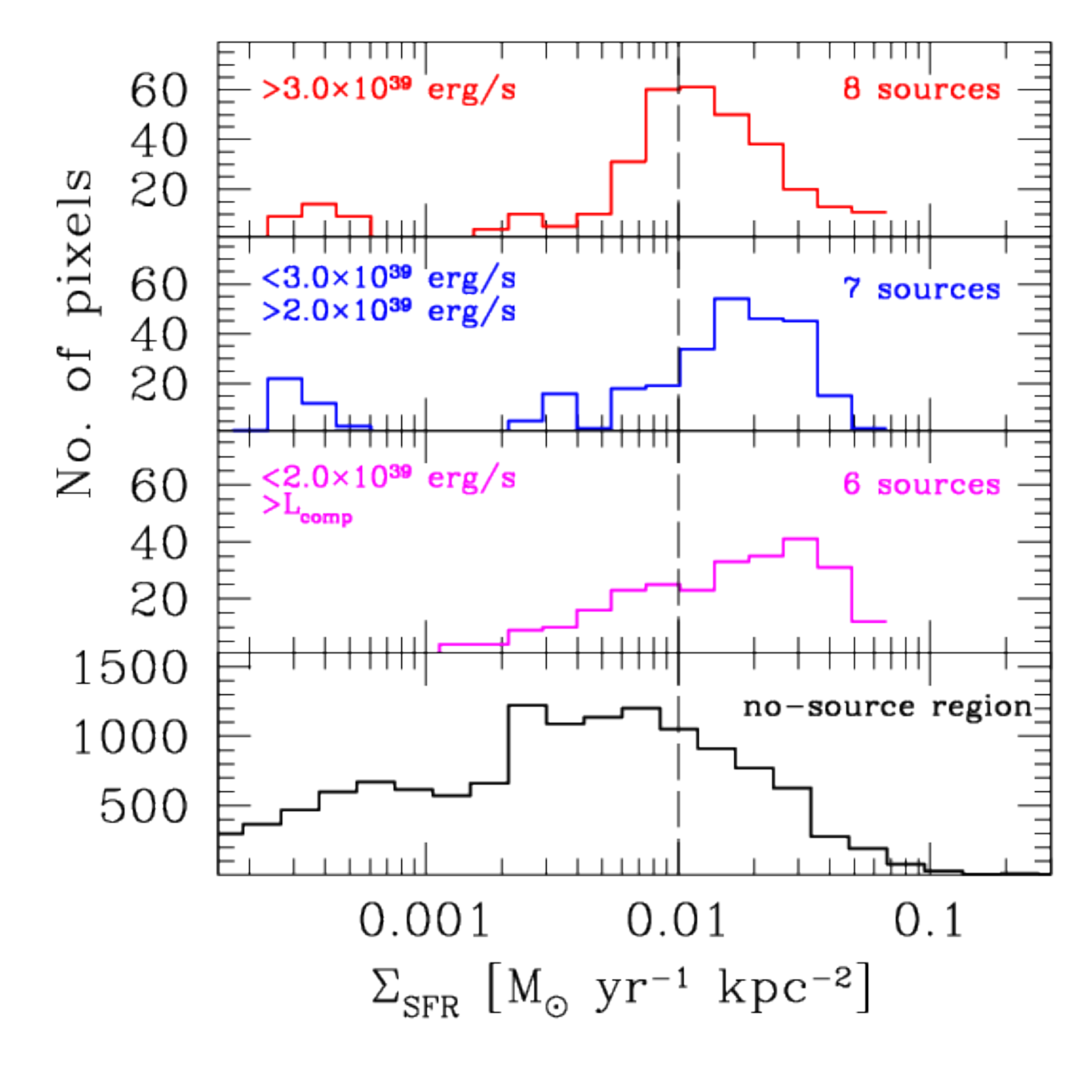}
\caption{Distribution of SFR density values surrounding the ULXs in NGC 2207/IC 2163. The SFR density was computed as described in Sect.~\ref{sec:sfr_distr} in the regions immediately surrounding the 21 individual X-ray sources in the colliding galaxy pair. From top to bottom, the histograms show the distribution of SFR density values around X-ray sources more luminous than $3\times 10^{39}\,\rm{erg}\,\rm{s}^{-1}$ (red), around sources with $2\times 10^{39}\,\rm{erg}\,\rm{s}^{-1} < L_{\rm{X}}\leq 3\times 10^{39}\,\rm{erg}\,\rm{s}^{-1}$ (blue), around sources with $L_{\rm{comp}} < L_{\rm{X}}\leq 2\times 10^{39}\,\rm{erg}\,\rm{s}^{-1}$ (magenta) and for all pixels where no sources were detected (black). To generate all four histograms we considered only points inside of the D25 contour of the galaxy pair. The histogram peaks for the three ULX groups are shifted slightly toward higher SFRs as the X-ray luminosity range decreases. The trend is driven by the FIR emission. In a more compelling way, the regions with no source detections peak at much lower values of SFR densities than where the X-ray sources are found.}
\label{fig:sfr_histo}
\end{center}
\end{figure}

We studied the distribution of the SFR densities around the detected ULXs, with the ultimate aim of constraining the ULX evolution in relation to the star formation time scale. The analysis that we discuss below is based on the SFR density image shown in Figure~\ref{fig:sfrmap}.

HMXBs can have large runaway velocities due to kicks caused by asymmetric explosions in the formation of the compact object. Since it has been suggested that ULXs might be bright HMXBs \citep{2003MNRAS.339..793G, 2012MNRAS.419.2095M}, high spatial velocities may also be relevant to ULXs.  A compact object with a high-mass companion can have an average runaway speed of the order of $\sim$$40\,\rm{km}\,\rm{s}^{-1}$, in the case of OB supergiant X-ray binaries or $\sim$$15\,\rm{km}\,\rm{s}^{-1}$ for Be/X-ray binaries \citep{1998A&A...330..201C}. \citet{2005MNRAS.358.1379C} measured a maximum average speed of $\sim$$30\,\rm{km}\,\rm{s}^{-1}$ based on the separation of star clusters and HMXBs in the Small Magellanic Cloud (SMC). Other studies found that the limiting average speed is $\sim$$50\,\rm{km}\,\rm{s}^{-1}$ \citep[e.g.,][]{2000A&A...364..563V, 2004MNRAS.348L..28K}. The average speed among those noted here, $\sim$$35\,\rm{km}\,\rm{s}^{-1}$, was adopted as a typical runaway velocity of our luminous X-ray sources. Assuming that the latter move at this speed over a $\sim$$30\,\rm{Myr}$ lifetime, appropriate for HMXBs and ULXs, they would be displaced by $\sim$$1\,\rm{kpc}$ at end of life with respect to their location at birth.  At the distance of NGC~2207/IC~2163 this corresponds to a maximum proper motion of $\sim$$5.5\arcsec$. The latter value was therefore used as the radius of each of the fiducial circular regions immediately surrounding the 21 high-luminosity X-ray sources in which to measure the level of local SFR activity. 

We divided the sources into three groups according to their X-ray luminosities: $L_{\rm{comp}} < L_{\rm{X}}\leq 2\times 10^{39}\,\rm{erg}\,\rm{s}^{-1}$, $2\times 10^{39}\,\rm{erg}\,\rm{s}^{-1} < L_{\rm{X}}\leq 3\times 10^{39}\,\rm{erg}\,\rm{s}^{-1}$, and $L_{\rm{X}} > 3\times 10^{39}\,\rm{erg}\,\rm{s}^{-1}$. The three groups include 6, 7 and 8 sources respectively. For each group, we created a corresponding SFR density image by selecting, from the original SFR density map, only the pixels which are covered by the fiducial circular regions centered on the sources in that luminosity group. In addition, we created a SFR density map for the area of the $D25$ ellipse where no sources were detected, by excluding the fiducial circular regions of all detected sources. Histograms of the values in these four images show the distributions of SFRs in the vicinities of the sources in each group and, to act as a control, in the regions away from all sources (see Fig.~\ref{fig:sfr_histo}).

The peaks of the histograms for the three ULX groups appear to shift slightly (by at most a factor of $\sim$2) toward {\em higher} SFR densities values as the X-ray luminosity range {\em decreases}. This is opposite to what one might have naively guessed. The slight trend is driven by the 24 $\mu$m emission rather than the FUV emission, i.e., by the younger ($\lesssim 10$ Myr, Calzetti et al. 2007; Rappaport et al. 2012, in preparation) star formation tracer of the two ingredients used to construct the SFR density image.  A vertical dashed line corresponding to the position of the peak in the histogram for the brightest sources is plotted in Fig.~\ref{fig:sfr_histo}. More in line with expectations, the distribution for the regions away from the sources peaks at much lower SFR densities. This shows that the fraction of pixels having high SFR density, $\sim$$0.01-0.03\,M_{\odot}\,\rm{yr}^{-1}\,\rm{kpc}^{-2}$, is much higher in the vicinity of bright X-ray sources, than in the field as a whole. This confirms our expectation that ULXs tend to be located close to star-forming regions.

To quantify the consistency, or lack thereof, among the three distributions (top three panels in Fig.~\ref{fig:sfr_histo}) we performed a KS test. The three KS D values are $D_{H,M}=0.18$, $D_{H,L}=0.24$ and $D_{M,L}=0.13$, respectively between the histograms for high and middle, high and low, middle and low luminosity groups shown in Fig.~\ref{fig:sfr_histo}. This yields two-sided KS statistics probabilities of $\sim$$3.1\times10^{-5}$, $\sim$$6.5\times10^{-8}$ and $\sim$$1.7\times10^{-2}$ respectively, indicating that the histograms are likely drawn from different underlying distributions. Note that the KS test tends to emphasize the region near the peak of the distribution, i.e., the region where the best Poisson statistics obtain.  Additionally, both systematic errors in estimating the SFR rate, as well as correlations in SFR among neighboring pixels, make the error estimates for these histograms difficult to evaluate.  With these caveats, and using the present data, we conclude that the ULX luminosity does not seem to depend strongly on the local SFR around the source, but further studies with larger samples will help clarify the situation.  There is little doubt, however, that there is a significant difference between the SFRs near ULXs and those from source-free regions.

Fig.~\ref{fig:sfr_histo} also reinforces the conclusion found in regard to the $N_{\rm{X}}-\rm{SFR}$ and $L_{\rm{X}}-\rm{SFR}$ relations, shown in Fig.~\ref{fig:nx_lx_sfr}, that while the {\em numbers} of ULXs are significantly correlated with the local SFR densities, the {\em luminosities} of those sources do not appear to be so correlated.

\section{Total X-ray luminosity per unit SFR}
\label{sec:tot_lx_sfr}

After having subtracted the predicted contribution from background AGNs, the collective X-ray luminosity of the compact source population with $L_{\rm{X}}>L_{\rm{comp}}$ within the $D25$ ellipse is $L_{\rm{XRB}}(0.5-8\,\rm{keV}) = (5.33\pm0.37)\times 10^{40}\,\rm{erg}\,\rm{s}^{-1}$. The total SFR integrated within the same region (Sect.~\ref{sec:int_sfr}) is $11.8 \,M_{\odot}\,\rm{yr}^{-1}$. 

We compare this result with the more extensive results of \citet{2012MNRAS.419.2095M}. First, we estimate the integrated SFR of NGC 2207/IC 2163 following the same method used in the latter paper, and that is based on the recipe of \citet{2006ApJS..164...38I}, which assumes a Salpeter IMF from 0.1 to 100 $M_{\odot}$: $\rm{SFR} = \rm{SFR_{NUV}^{0}}+\rm{SFR_{IR}}$. $\rm{NUV}^{0}$ is the near-ultraviolet (2312\,\AA) luminosity uncorrected for dust attenuation and IR is the $8-1000\,\mu \rm{m}$ luminosity \citep[see Sect.~5 of][for details]{2012MNRAS.419.2095M}. We obtain a total SFR of $23.7\,M_{\odot}\,\rm{yr}^{-1}$, similar to that obtained in Sect.~\ref{sec:int_sfr} under the same IMF assumption (cf. 18.8 $M_\odot$ yr$^{-1}$; SFR estimations usually have uncertainties of $\sim$50\%). In general FUV and NUV emissions are consistent with each other within $5-7\%$, so the main difference between the relation above and the integrated SFR measured as described in  \S\ref{sec:int_sfr}, is due to the different IR estimator.

Using the calibration from \citet{2012MNRAS.419.2095M}, (their eq.~(22)) we find that the above SFR implies that $L_{\rm{XRB}}(>10^{36}\,\rm{erg}\,\rm{s}^{-1}) = 23.7\cdot 2.6\times 10^{39} = 6.2\times 10^{40}\,\rm{erg}\,\rm{s}^{-1}$. Taking into account the difference in the luminosity limit, by integrating the HMXB luminosity distribution from the same paper, we obtain $L_{\rm{XRB}}(>L_{\rm{comp}})= 4.8\times 10^{40}\,\rm{erg}\,\rm{s}^{-1}$. The total X-ray luminosity measured for NGC~2207/IC~2163 is in agreement with the latter result to within $\sim$10\%. Analogously, we can use the same SFR, in eq.~(20) of \citet{2012MNRAS.419.2095M}, to predict that $N_{\rm{XRB}}(L>L_{\rm{comp}}) = 17.9$, for the same luminosity threshold. Comparing this with the observed number of sources, 19.3, the agreement is rather good (the difference between the solid line and the points in Fig. \ref{fig:nx_lx_sfr} is much larger, due to the different method of calculating the SFR). This supports the suggestive results by \citet{2003MNRAS.339..793G} that the ULX population -- at least at the bottom end of their luminosity range -- might be an extension of the HMXB population at high luminosities.

\section{Discussion}
\label{sec:discussion}

In Figures~\ref{fig:montage}, \ref{fig:hst_src} and \ref{fig:sfrmap} it appears that IC~2163 is forming stars more actively than NGC~2207 (note the much more extensive dust and gas content in the former), although the latter galaxy hosts most of the detected ULXs. A comparison between UV and $24\,\mu \rm{m}$ emissions (Fig. \ref{fig:montage}) shows that the latter is more enhanced, suggesting that the star formation activity in IC~2163 may be more recent ($< 10$ Myr). This might indicate a possible age effect in the distribution of ULXs \citep[see e.g., \S7.2 and \S7.3 in][]{2005A&A...431..597S}. The specific SFR, i.e., the SFR per unit stellar mass, could help us in investigating the latter hypothesis. We measured the integrated SFR and stellar mass of the two galaxies separately. The area of the sky covered by IC~2163 was defined by visually inspecting its optical and infrared images. We did not use the $D25$ ellipse as it includes a large fraction of the nearby NGC~2207.
The SFR was estimated as described in Sect. \ref{sec:tot_lx_sfr} and the stellar mass using the $K_{S}$-band image from the 2MASS Large Galaxy Atlas (LGA)\footnote{http://irsa.ipac.caltech.edu/applications/2MASS/LGA/}, assuming an effective $B-V$ color of 0.67 mag (provided by HyperLeda\footnote{http://leda.univ-lyon1.fr/}) and using the calibration from \citet{2001ApJ...550..212B}.

The integrated SFR of IC~2163 is $9.1\,M_{\odot}\,\rm{yr}^{-1}$ and its inferred stellar mass is $5.2\times 10^{10}\,M_{\odot}$. Five ULXs out of 21 are located within its area. This yields a $N_{\rm{X}}/\rm{SFR}\simeq 0.55\pm0.25$ ($M_{\odot}$/yr)$^{-1}$ and specific star formation rate $\rm{SFR}/\rm{M_{\star}}\simeq 1.75 \times 10^{-10}\,\rm{yr}^{-1}$.
Subtracting these values from those measured within the entire $D25$ ellipse of the galaxy pair, we estimated for NGC~2207 a total SFR of $14.6\,M_{\odot}\,\rm{yr}^{-1}$ and stellar mass $1.2\times 10^{11}\,M_{\odot}$. This yields a $\rm{SFR}/\rm{M_{\star}}\sim 1.22\times 10^{-10}\,\rm{yr}^{-1}$ and $N_{\rm{X}}/\rm{SFR}\sim 1.09\pm0.27$ ($M_{\odot}$/yr)$^{-1}$. The difference in specific SFR is not large. The difference in $N_{\rm{X}}/\rm{SFR}$ is only marginally significant. The source number in IC~2163 is a factor of 1.4 lower than that predicted by the calibration from \citet{2012MNRAS.419.2095M}, while in the rest of the merging system it is a factor of 1.5 higher than predicted.

We can compare these properties with those of the Antennae (NGC~4038/39). Using the same method as above to estimate the integrated SFR and stellar mass, \citet{2012MNRAS.419.2095M} obtain $\rm{SFR}=5.4\,M_{\odot}\,\rm{yr}^{-1}$ and $\rm{M_{\star}} = 3.1\times 10^{10}\, M_{\odot}$. Within the same area of sky the same authors detected 5 ULXs. This corresponds to only somewhat larger numbers of ULXs per unit mass and per unit SFR in comparison with those of NGC~2207/IC~2163. 

\begin{deluxetable*}{cccccc}
\tablewidth{0pt}
\tabletypesize{\scriptsize}
\tablecaption{\label{tab:sfr} Summary of Star Formation Rates}
\tablehead{
	\colhead{Galaxy} &
	\colhead{$L_{\rm{X}}^{\rm tot}$} &
	\colhead{$M_*$}  & 
	\colhead{SFR} &
	\colhead{SFR/$M_*$} &
	\colhead{$N_{\rm{X}}$/SFR} \\
	\colhead{} & 
	\colhead{($10^{40}$ ergs s$^{-1})$} &
	\colhead{($10^{11} M_\odot$)} &
	\colhead{($M_\odot$ yr$^{-1}$)} & 
      	\colhead{($10^{-10}$ yr$^{-1}$)} &
	\colhead{(yr $M_\odot^{-1}$)} \\
	(1)  & (2) & (3) & (4) & (5) & (6)
}
\startdata 
NGC 2207 & 4.3 $\pm$ 0.3 & 1.2 & 14.6 & 1.22 & 1.09 $\pm$ 0.27 \\
IC 2163 & 1.0 $\pm$ 0.2 & 0.52 & 9.1 & 1.75 & 0.55 $\pm$ 0.25 \\
Entire $D_{25}$ &  5.3 $\pm$ 0.4 & 1.7 & 23.7 & 1.39 & 0.89 \\
Antennae & 1.7 & 0.31 & 5.4 & 1.74 & 0.93
\enddata
\tablecomments{(1) Galaxy; (2) collective 0.5--8 keV luminosity of the
  compact source population with with
  $L_{\rm{X}} > 10^{39}$ ergs s$^{-1}$; (3) mass in stars; (4) total
  star formation rate, $\rm{SFR}= 4.6 \times 10^{-44} L_{\rm{IR}} +
  1.2 \times 10^{-43} L_{\rm{NUV,obs}}$ as described in \S 7; (5)
  specific star formation rate; (6) number of luminous X-ray sources
  per unit SFR. }
\end{deluxetable*}

The above results for galaxy masses, SFRs, specific SFRs, and luminous X-ray sources per SFR are summarized in Table \ref{tab:sfr}.

We tried to identify possible counterparts to the 21 ULXs hosted by NGC~2207/IC~2163. We cross-matched their locations with the Naval Observatory Merged Astrometric Dataset (NOMAD)\footnote{http://www.usno.navy.mil/USNO/astrometry/optical-IR-prod/nomad} \citep{2004AAS...205.4815Z}. This catalog is a merger of data from the Hipparcos, Tycho-2, UCAC-2 and USNO-B1 catalogs, supplemented by photometric information from the Two Micron All Sky Survey (2MASS) final release point source catalog. We used a match radius of $2\arcsec$ bearing in mind that the $99\%$ uncertainty circle of {\em Chandra} absolute positions for sources within $3\arcmin$ of the aimpoint is $0.8\arcsec$. The choice of $2\arcsec$ match radius is rather conservative considering that the typical mean error on coordinates in the NOMAD catalog is also $\lesssim 0.8 \arcsec$. 
We found possible counterparts in the USNO B1.0 catalog for 7 ULXs, i.e., source  nos. 2, 4, 7, 14, 15, 25, and 28. For all these objects, the catalog provides, among other measurements, the B- and R-band magnitudes and the proper motions. All the optical objects are separated from the corresponding Chandra ULX positions by angular distances ranging from $1.2\arcsec$ (for source \#14) to $2.0\arcsec$ (for source no. 4). Based on the accuracy of the absolute Chandra positions, this indicates that there are no compelling candidate counterparts among the nearly matched sources. Moreover, sources 2, 4 and 15 have measured proper motions, which indicate that these are most probably foreground objects. A more accurate search of the ULX counterparts would require the use of HST data, which are publicly available in the HST archive. A forthcoming paper will include such a study.

The search for optical counterparts can also be useful to identify possible background AGNs among our ULX sample. In particular, sources 4, 16, 21 and 27 are located outside the spiral structures of the two galaxies. For the latter three there is no counterpart in NOMAD. As mentioned above, one match was found for source 4, with a separation of $\approx 2.0\arcsec$ from its {\em Chandra} coordinates. This source is the brightest X-ray object detected in the vicinity of NGC 2207/IC 2163. The R-band magnitude of the optical object is $14.7$. This can be converted into a flux $f_{\rm{R}}\simeq 1.7\times 10^{-11} \,\rm{erg}\,\rm{cm}^{-2}\,\rm{s}^{-1}$ which yields, in turn, $\log(f_{\rm{0.5-8 keV}}/f_{\rm{R}})\simeq -2.4$. According to \citet{2004AJ....128.2048B} this object should then not be an AGN. This is further supported by the fact that the hardness ratio of source 4 differs from that of the central AGN in NGC~2207 (see Table \ref{tab:xray}). However, we note that  low luminosity AGN populate the same region of the X-ray flux vs $R$-band magnitude plane \citep[Fig. 7 in][]{2004AJ....128.2048B} as our source \#4. On statistical grounds, the predicted number of background sources above the luminosity of source 4 is only 0.09. For a Poisson distribution, the probability that the brightest ULX in our galaxy pair is a background AGN is only $8\%$. Assuming that the observed XLF of the ULXs extends to the highest luminosities with the same slope, the predicted number of X-ray sources above the same threshold would be 1. Thus, a ULX interpretation for source \#4 is the most likely.

\section{Summary and conclusions}

We have introduced a new technique to investigate the spatial and luminosity distributions of X-ray binaries in star forming galaxies as a function of the {\em local} SFRs. We have applied this technique to study the population of 21 ULXs in the colliding galaxy pair NGC~2207/IC~2163. This is comparable with the largest number of ULXs per unit mass in any galaxy that we know of, in particular the Antennae.

Using the prescription by \citet{2008AJ....136.2782L}, we constructed an image of SFR density, in units of $M_{\odot}\,\rm{yr}^{-1}\,\rm{kpc}^{-2}$, of NGC~2207/IC~2163, by combining the {\em Galex} FUV and {\em Spitzer} 24-$\mu$m images. We find that the global relation between the cumulative number of X-ray point sources and the integrated SFR of the host galaxy also holds on {\em local} scales. We were not able to investigate in detail the corresponding local $L_{\rm{X}}-\rm{SFR}$ relation due to the small number of X-ray sources. 

We studied the distribution of the SFR density around the detected ULXs, with the ultimate aim of constraining the ULX evolution in relation to the star formation time scale. We find that the peaks in the SFR density distributions around ULXs appear to shift slightly toward {\em higher} SFR density values as the X-ray luminosity range {\em decreases}. The regions with no source detections, however, do peak at much lower SFR densities than where the X-ray sources are found. The fraction of pixels having high SFR density, $\sim$$0.01-0.03\,M_{\odot}\,\rm{yr}^{-1}\,\rm{kpc}^{-2}$, is higher in the vicinity of bright X-ray sources, than in the field. 

We find that the number and luminosity of ULXs per unit SFR are in agreement, to within 10\%, with those predicted by the global relations of \citet{2012MNRAS.419.2095M}. This supports the suggestion of \citet{2003MNRAS.339..793G} that the lower luminosity end of the ULX population (e.g., $1-3 \times 10^{39}$ ers s$^{-1}$) might be an extension of the HMXB population to higher luminosities. 

We attempted to investigate possible age effects in the distribution of ULXs across the galaxy pair NGC~2207/IC~2163. We find that the difference in specific SFR ($\rm{SFR}/\rm{M_{\star}}$) between NGC~2207 and IC~2163 is not large. The difference in $N_{\rm{X}}/\rm{SFR}$ is only marginally significant. The source number in IC~2163 is a factor of 1.4 lower than that predicted by the calibration from \citet{2012MNRAS.419.2095M}, while in the rest of the merging system it is a factor of 1.5 higher than predicted. 

We tried to identify possible counterparts to the 21 ULXs hosted by NGC~2207/IC~2163.  In particular, our search of the USNO-B1 catalog emphasized possible optical counterparts to the brightest detected source (no. 4), as well as for sources 16, 21 and 27, which are located {\em outside} the spiral structures of the two galaxies.  In general, the relatively large separations (i.e., $\gtrsim 1''$) between possibly interesting stellar images and the {\em Chandra} coordinates suggest that there are no compelling candidate counterparts.  The closest optical object to the brightest detected source (no. 4) found in the USNO-B1 catalog does not have the properties of a background AGN. A more definitive search for ULX counterparts would require the use of HST data, which is publicly available in the HST archive. We plan to pursue this in a forthcoming paper.

{\em Note Added to the Manuscript:}  After this work was complete, we learned from 
M. Kaufman (2013, private communication) of some useful comparisons with 
their earlier {\em XMM-Newton} observations of NGC 2207/IC 2163 (Kaufman et 
al.~2012).  In particular {\em Chandra} source \#11 apparently coincides with a
variable, nonthermal radio source, and is therefore more likely to be a background 
AGN than a ULX.  The soft, extended {\em Chandra} X-ray source \#28 is close to 
the radio core of {\em feature i} in Kaufman et al.~(2012), and the extended emission is 
likely to come from this starburst region.

\acknowledgments
The authors are grateful to Michele Kaufman, Mark Krumholz, Adam Leroy and Pepi Fabbiano for helpful discussions. The authors thank the anonymous referee for helpful comments that improved this paper. SR, DP, and AL acknowledge support from {\em Chandra} Grants GO1-12111X and GO2-13105A. We made use of \textit{Chandra} archival data and software provided by the \textit{Chandra} X-ray Center (CXC) in the application package CIAO. We also utilized the software tool SAOImage DS9, developed by Smithsonian Astrophysical Observatory. The FUV, 3.6 $\mu$m, and 24 $\mu$m images were taken from {\em Galex} and {\em Spitzer} archives, respectively. The \textit{Spitzer Space Telescope} is operated by the Jet Propulsion Laboratory, California Institute of Technology, under contract with the NASA. \textit{GALEX} is a NASA Small Explorer, launched in 2003 April. We also made use of data products from the Two Micron All Sky Survey (2MASS), which is a joint project of the University of Massachusetts and the Infrared Processing and Analysis Center/California Institute of Technology, funded by NASA and the National Science Foundation. Helpful information was found in the NASA/IPAC Extragalactic Database (NED) which is operated by the Jet Propulsion Laboratory, California Institute of Technology, under contract with the National Aeronautics and Space Administration.

\end{document}